\documentclass[a4paper,11pt]{article}
\pdfoutput=1 
\usepackage{jheppub} 
\usepackage{amsmath,amsfonts,amssymb,mathrsfs,graphicx,color,longtable,bm,wasysym}
\usepackage{hyperref}
\usepackage{graphicx}
\usepackage{array}
\usepackage{color}
\usepackage[usenames,dvipsnames,table]{xcolor}
\usepackage{tikz}
\usepackage[utf8]{inputenc}
\usepackage{epsfig,relsize,xspace}
\usepackage[T1]{fontenc}
\usepackage{bbold}
\usepackage{float}
\usepackage{amsmath}
\usepackage{empheq}
\usepackage{booktabs}
\usepackage{color}
\usepackage[utf8]{inputenc}
\usepackage{xspace}
\usepackage{scalerel}
\usepackage[most]{tcolorbox}
\usepackage{multirow}
\usepackage{scalefnt}
\usepackage{bold-extra}
\usepackage[shortlabels]{enumitem}
\usepackage[tikz]{bclogo}
\usepackage{subcaption}
\usepackage{tikz-feynman}
\usepackage{cancel}
\usepackage{verbatim}
\usetikzlibrary{arrows,shapes}
\usepackage{afterpage}
\usepackage{graphicx,grffile}
\usepackage{slashed}
\usepackage{soul}

\allowdisplaybreaks
 
\definecolor{nicered}{rgb}{0.7,0.1,0.1}
\definecolor{nicegreen}{rgb}{0.1,0.5,0.1}
\definecolor{niceblue}{rgb}{0.0,0.1,0.7}
\hypersetup{colorlinks,citecolor=niceblue,linkcolor=niceblue,urlcolor=niceblue}
                                      
\def\bm#1{\mbox{\boldmath$#1$\unboldmath}}
\def \beq{\begin{equation}}
\def \eeq{\end{equation}}
\def \bea{\begin{eqnarray}}
\def \eea{\end{eqnarray}}

\begin{document}

\def\arraystretch{1.25}

\title{\boldmath  Off-shell Higgs production at the LHC \\ as a probe of the trilinear Higgs coupling  }

\author[1]{Ulrich Haisch}

\author[1]{and Gabri{\"e}l Koole}

\affiliation[1]{Max Planck Institute for Physics, F{\"o}hringer Ring 6,  80805 M{\"u}nchen, Germany}

\emailAdd{haisch@mpp.mpg.de}
\emailAdd{koole@mpp.mpg.de}

\abstract{In the context of the Standard Model effective field theory~(SMEFT) we examine the constraints on the trilinear Higgs coupling that originate from off-shell Higgs production in proton-proton collisions. Our calculation of the $gg \to h^\ast \to Z Z \to 4 \ell$ process includes   two-loop corrections to gluon-gluon-fusion Higgs production and  one-loop corrections to the Higgs propagator and its decay. Employing a matrix-element based kinematic discriminant we determine the reach of LHC~Run~3 and the high-luminosity option of the LHC in constraining the relevant SMEFT Wilson coefficients. We present constraints that are not only competitive with but also complementary to the projected indirect limits that one expects to obtain from inclusive measurements of single-Higgs production processes at future LHC runs.  }

\maketitle

\section{Introduction}
\label{secf:intro}

In the pursuit of observing or constraining deviations from the Standard Model~(SM) picture, measuring the couplings of the  Higgs boson to the other bosons, fermions and itself lies at the heart of the physics programme at the Large Hadron Collider (LHC). The couplings of the  scalar resonance to electroweak~(EW) gauge bosons and third-generation fermions have been tested extensively, showing agreement with those of the SM Higgs boson at the level of $(10 - 20)\%$~\cite{CMS-PAS-HIG-19-005,ATLAS:2020qdt}. The strength of these couplings will be further scrutinised at the high-luminosity option of the LHC~(HL-LHC) and possible future colliders. In contrast to the couplings of the  Higgs boson to EW gauge bosons and third-generation fermions, the Higgs couplings to light matter fields  and its self-interactions are largely unexplored at present (cf.~for example~\cite{Haisch:2017bpz} for an overview). 

Within the SM, the trilinear and quartic couplings of the physical Higgs field $h$ are parametrised by the scalar potential
\begin{align} \label{eq:VSM}
V_{\rm SM} = \frac{m_h^2}{2}  \hspace{.25mm} h^2 +  \lambda v h^3 + \frac{\kappa}{4}  \hspace{.25mm} h^4 \ ,
\end{align}
where $m_h \simeq 125 \, {\rm GeV}$ and $v \simeq 246 \, {\rm  GeV}$ are the Higgs mass and vacuum expectation value~(VEV), respectively, and the trilinear ($\lambda$) and quartic ($\kappa$) Higgs couplings obey the relation
\begin{align}\label{eq:smcouprel}
\lambda = \kappa = \frac{m_h^2}{2 \hspace{.125mm} v^2} \simeq 0.13 \ .
\end{align}
The SM potential is thus fully determined by only two parameters, namely $v$ and $\lambda$. However, many beyond the SM (BSM) scenarios allow for deviations of the Higgs self-couplings with respect to their SM predictions (a comprehensive collection of such theories can be found in the white paper~\cite{DiMicco:2019ngk}) and, consequently, could imply a departure from the SM~relation~(\ref{eq:smcouprel}). Measuring or constraining the Higgs  self-couplings is therefore essential to our understanding of the mechanism of electroweak symmetry breaking~(EWSB) and furthermore provides a way to probe the existence of new physics.

The aforementioned relatively loose constraints on the Higgs self-interactions are due to the smallness of the cross sections of double-Higgs and triple-Higgs production, the go-to observables for testing the Higgs potential because of their direct sensitivity to $\lambda$ and $\kappa$. As the triple-Higgs cross section is $\mathcal{O}(0.1 \text{ fb})$ for proton-proton~($pp$) collisions at a centre-of-mass energy of $\sqrt{s} = 14 \, {\rm TeV}$ even at the end of HL-LHC, having an integrated luminosity of $3 \, {\rm ab}^{-1}$ at hand, the quartic Higgs coupling will remain unexplored (cf.~\cite{Papaefstathiou:2015paa,Chen:2015gva,Fuks:2015hna,Kilian:2017nio,Fuks:2017zkg,Liu:2018peg,Bizon:2018syu,Borowka:2018pxx,Papaefstathiou:2019ofh,Chiesa:2020awd} for the prospects to  determine the quartic Higgs coupling at future colliders). The  HL-LHC prospects for double-Higgs production are considerably better but still remain challenging as the cross section is  $\mathcal{O}(33 \, {\rm fb})$ in $pp$ collisions at $\sqrt{s} = 14 \, {\rm TeV}$. As a result, only~${\cal O} (1)$ determinations of the trilinear Higgs coupling from double-Higgs production seem to be possible at the LHC~---~see~\cite{ATL-PHYS-PUB-2018-053,CMS-PAS-FTR-18-019} for  the latest prospect studies by ATLAS and CMS. With this in mind, other methods of constraining the trilinear Higgs coupling have been proposed in recent years. The indirect approach first outlined in~\cite{McCullough:2013rea}, where the sensitivity to $\lambda$ arises from loop corrections to the process $e^+ e^- \to Zh$, was later extended to the Higgs production and decay modes relevant at the LHC~\cite{Gorbahn:2016uoy,Degrassi:2016wml,Bizon:2016wgr,Maltoni:2017ims,Gorbahn:2019lwq,Degrassi:2019yix}, at lepton colliders~\cite{DiVita:2017vrr,Maltoni:2018ttu} and to EW precision observables~\cite{Degrassi:2017ucl,Kribs:2017znd}. 

Analyses of the constraints on the trilinear Higgs coupling including indirect probes have been presented in~\cite{DiVita:2017eyz,Maltoni:2017ims,Rossi:2020xzx,Degrassi:2021uik} and recently also by ATLAS and CMS~\cite{CMS-PAS-FTR-18-020,ATL-PHYS-PUB-2019-009,ATLAS-CONF-2019-049,CMS-PAS-HIG-19-005}. Generally, the direct constraints on $\lambda$ obtained through double-Higgs production were shown to furnish the most stringent bounds, but indirect constraints from single-Higgs production processes have the potential to be competitive or could fulfil at least a complementary role. As already pointed out in the works~\cite{Degrassi:2016wml,Bizon:2016wgr,Haisch:2017bpz}, including measurements of differential distributions of single-Higgs processes could turn out to be crucial due to their non-trivial dependence on $\lambda$. This~point was further investigated in the article~\cite{Maltoni:2017ims}, where it was found that the associated production of the  Higgs together with a EW gauge boson~($Vh$) or a top-antitop pair~($t \bar{t} h$) provide additional sensitivity to $\lambda$ at the differential level. 

In~this work, we investigate the LHC~Run~3 and the HL-LHC sensitivity to modifications of the trilinear Higgs coupling from off-shell Higgs production in the gluon-gluon fusion~(ggF) channel, considering  decays to  four charged leptons, i.e.~$h^\ast \to Z Z \to 4 \ell$. Within the context of the SM effective field theory~(SMEFT), we study the effects of a modified trilinear Higgs coupling through a shape analysis of differential distributions. In particular, the use of a matrix-element based kinematic discriminant to improve the sensitivity in regions where the Higgs boson is off-shell will allow us to put constraints on the trilinear Higgs coupling that are not only competitive with but also complementary to constraints from inclusive single-Higgs production cross-section measurements  at future~LHC~runs.

This article is organised as follows. In Section~\ref{sec:bsmpar}, we the introduce the effective interactions that are relevant for the computations performed in our paper. The calculations of the loop corrections to the production, propagator and decay of the Higgs boson in the process $gg \to h^\ast \to Z Z \to 4 \ell$ are presented in Section~\ref{sec:BSMcorrs}. Our numerical analysis is performed in Section~\ref{sec:numan}, where we also present our LHC~Run~3 and HL-LHC projections for the constraints on the trilinear Higgs coupling  and compare them to previous results. We conclude and present an outlook in Section~\ref{sec:concl}. Additional material is relegated to Appendix~\ref{app:higgswidth} and~Appendix~\ref{app:HLLHCprojections}. 

\section{Parametrisation of BSM effects}
\label{sec:bsmpar}

To allow for a  model-independent analysis, we work  in  the framework of the SMEFT~\cite{Buchmuller:1985jz,Grzadkowski:2010es,Brivio:2017vri} to parametrise the possible BSM physics entering the trilinear Higgs coupling. While efforts are being made to obtain a general fit of the many dimension-six SMEFT operators  to the available experimental data (see~for~example~\cite{Ellis:2018gqa,Ethier:2021bye} for some recent results), requiring absolute model-independence can in practice impede the extraction of meaningful results due to the large number of free parameters in the fit. In~this article, we therefore consider only the subset of dimension-six CP-even operators in the so-called Strongly Interacting Light Higgs~(SILH) basis~\cite{Giudice:2007fh} that are build purely from SM Higgs doublets $H$ and derivatives:
\begin{align}\label{eq:operatorsd6}
\mathcal{O}_{6} =  - \lambda \left  | H   \right |^6 \, , \qquad 
\mathcal{O}_{H} =  \frac{1}{2} \hspace{0.25mm} \Big( \partial_{\mu} |H|^2 \Big)^2 \, , \qquad 
\mathcal{O}_{T} =   \frac{1}{2} \hspace{0.25mm} \Big( H^{\dagger} \! \stackrel{\leftrightarrow}{D_{\mu}} \!   H \Big)^2 \, .  
\end{align}
Here  we have used the short-hand notation $H^{\dagger} \! \stackrel{\leftrightarrow}{D_{\mu}}\! H =  H^{\dagger}  D_{\mu} H - (D_{\mu}H)^{\dagger} H$. We thus implicitly assume that the couplings of the Higgs boson to EW gauge boson and fermions are SM-like and focus our attention on the self-interactions contained in~(\ref{eq:VSM}). Note that the redundant operators $\mathcal{O}_R=  |H|^2  \hspace{0.25mm} |D_{\mu} H|^2$ and $\mathcal{O}_D=  |D^2 H|^2$ do not appear in~(\ref{eq:operatorsd6}) as they can be removed in the full SILH basis via an appropriate redefinition of the Higgs field or equivalently its equations of motion~\cite{Elias-Miro:2013eta}. Furthermore, as the operator $\mathcal{O}_T$ does not modify the trilinear Higgs coupling and is moreover severely constrained through measurements of the $\rho$ parameter describing the degree of custodial symmetry violation~\cite{ParticleDataGroup:2020ssz}, it is irrelevant for the purpose of this work. 

Within these restrictions, it is sufficient to supplement the SM Lagrangian with the effective operators $\mathcal{O}_{6}$ and $\mathcal{O}_{H}$ only, resulting in the following effective Lagrangian
\begin{align}\label{eq:lagrangian}
\mathcal{L} = \mathcal{L}_{\rm SM} + \sum_{i = 6,H} \frac{\bar{c}_i}{v^2} \hspace{0.5mm} \mathcal{O}_i \ ,
\end{align}
where $\bar{c}_i$ are  Wilson coefficients understood to be evaluated at the EW scale,~i.e.~$\mu = {\cal O} (v)$. Upon EWSB  and canonical normalisation of the Higgs kinetic term, one finds that the contributions from~(\ref{eq:lagrangian}) 
that are relevant for our article can be written as 
\begin{align}\label{eq:potd6}
{\cal L} \supset  -\lambda c_3 \hspace{0.25mm} v  h^3 \, , 
\end{align}
where
\begin{align} \label{eq:c3}
c_3 = 1+\bar{c}_6 - \frac{3 }{2} \hspace{0.5mm} \bar{c}_H \,, 
\end{align}
Note that while the sensitivity to the Higgs trilinear coupling in the process $gg \to h^* \to Z Z \to 4 \ell$ arises through next-to-leading order~(NLO) EW corrections, the Wilson coefficient~$\bar{c}_H$ appears already  at the Born level as it not only modifies the Higgs trilinear coupling, but also causes a universal shift of all couplings of the  Higgs boson. We~therefore include~$\bar{c}_H$ at leading order~(LO) in our numerical analysis (Section~\ref{sec:onshcomp}), while parametrising the NLO corrections involving insertions of the Higgs trilinear coupling (Section~\ref{sec:BSMcorrs}) solely  via the Wilson coefficient $\bar{c}_6$. Including $\bar{c}_H$ also at NLO would require taking into account all SM~NLO~EW effects, but this has a negligible effect on the extraction of $\bar{c}_6$~\cite{Maltoni:2017ims,Maltoni:2018ttu}. Effects coming from higher-dimensional pure Higgs operators such as ${\cal O}_8 = -\lambda \bar{c}_8/v^4 \left |H \right |^4$ are not considered, but could in principle be implemented by shifting the correction factor in~(\ref{eq:c3}) in an appropriate manner, for example by $2 \hspace{0.25mm} \bar{c}_8$ in the case of ${\cal O}_8$.

\section{Description of the $\bm{ gg \to h^\ast \to Z Z}$ calculation}
\label{sec:BSMcorrs}

In this section we briefly describe  the calculation of the $\mathcal{O}(\lambda)$ corrections to the process $gg \to h^\ast \to Z Z$ that arise in the context of~(\ref{eq:lagrangian}). The corrections associated to insertions of the operator ${\cal O}_6$ are illustrated in Figure~\ref{fig:diagrams}. They fall into three classes: (i) corrections to ggF Higgs production, (ii) corrections to the Higgs propagator and (iii) corrections to the Higgs decay. In the following, we will discuss separately each of these three ingredients as well as their implementation in our Monte Carlo~(MC) code. 

\subsection{Higgs production}

\begin{figure}[!t]
\centering
\begin{subfigure}[c]{\linewidth}
\centering
 \begin{tikzpicture}
         \begin{feynman}
        \vertex (a1)  {\(g\)};  
        \vertex[right=1.5cm of a1] (a2);
        \vertex[right=1.5cm of a2] (a3);
        \vertex[right=1.15cm of a3] (a4);
        \vertex[below=1.7cm of a1] (b1)  {\(g\)};
        \vertex[right=1.5cm of b1] (b2);
        \vertex[right=1.5cm of b2] (b3);
        \vertex[right=1.15cm of b3] (b4);
        \vertex[below=0.85cm of a4] (c1);
        \vertex[right=1.5cm of c1] (c2);
        \vertex[right=1cm of c2] (c3);
        \vertex[above=.8cm of c3] (c4){\(Z\)};
        \vertex[below=.8cm of c3] (c5){\(Z\)};
       \diagram* {
        (a1) -- [gluon, thick] (a2);
        (b1) -- [gluon, thick] (b2);
        (a2) -- [fermion,  edge label=\(t\), thick] (a3) -- [fermion,  edge label'=\(t\), thick] (b3) -- [fermion,  edge label=\(t\), thick] (b2) -- [fermion,  edge label'=\(t\), thick] (a2);
        (a3) -- [scalar, edge label=\(h\), thick] (c1) -- [scalar, edge label=\(h\), thick] (c2);
        (b3) -- [scalar, edge label'=\(h\), thick] (c1);
        (c2) -- [boson, thick] (c4);
        (c2) -- [boson, thick] (c5);
         };
         \filldraw[color=black, fill=black] (a2) circle (0.04);
         \filldraw[color=black, fill=black] (a3) circle (0.04);
         \filldraw[color=black, fill=black] (b2) circle (0.04);
         \filldraw[color=black, fill=black] (b3) circle (0.04);
         \filldraw[color=black, fill=black] (c2) circle (0.04);
         \node[square dot,fill=black,large] (d) at (c1){};
      \end{feynman}
\end{tikzpicture}
\end{subfigure}

\vspace{8mm}

\begin{subfigure}[c]{\linewidth}
\centering
\begin{tikzpicture}
         \begin{feynman}
        \vertex (a1)  {\(g\)};  
        \vertex[right=1.5cm of a1] (a2);
        \vertex[below=1.7cm of a1] (b1)  {\(g\)};
        \vertex[right=1.5cm of b1] (b2);
        \vertex[below=0.85cm of a2] (c1);
        \vertex[right=1.2cm of c1] (c2);
        \vertex[right=0.925cm of c2] (h2);
        \vertex[right=1.2cm of h2] (h3);
        \vertex[right=0.825cm of h3] (h4);
        \vertex[right=1cm of h4] (c3);
        \vertex[above=.8cm of c3] (c4) {\(Z\)};
        \vertex[below=.8cm of c3] (c5) {\(Z\)};
        
       \diagram* {
        (a1) -- [gluon, thick] (a2);
        (b1) -- [gluon, thick] (b2);
        
        (a2) -- [fermion, thick, edge label=\(t\)] (b2) -- [fermion, thick, edge label'=\(t\)] (c2) -- [fermion, thick, edge label'=\(t\)] (a2);
        
        (c2) -- [scalar, thick,  edge label=\(h\)] (h2);
        (h2) -- [scalar, half right, looseness=1.8, edge label'=\(h\), thick] (h3);
        (h2) -- [scalar, half left, looseness=1.8, edge label=\(h\), thick] (h3);
        (h3) -- [scalar, thick,  edge label=\(h\)] (h4);
        
        (h4) -- [boson, thick] (c4);
        (h4) -- [boson, thick] (c5);
         };
         \filldraw[color=black, fill=black] (a2) circle (0.04);
         \filldraw[color=black, fill=black] (b2) circle (0.04);
        \filldraw[color=black, fill=black] (c2) circle (0.04);
         \filldraw[color=black, fill=black] (h4) circle (0.04);
         \node[square dot,fill=black,large] (d) at (h2){};
         \node[square dot,fill=black,large] (d) at (h3){};
      \end{feynman}
\end{tikzpicture}
\end{subfigure}

\vspace{9mm}

\centering
\begin{subfigure}[c]{\linewidth}
\centering
 \begin{tikzpicture}
         \begin{feynman}
        \vertex (a1)  {\(g\)};  
        \vertex[right=1.5cm of a1] (a2);
        \vertex[below=1.7cm of a1] (b1)  {\(g\)};
        \vertex[right=1.5cm of b1] (b2);
        \vertex[below=0.85cm of a2] (c1);
        \vertex[right=1.2cm of c1] (c2);
        \vertex[right=1.1cm of c2] (h2);
        \vertex[right=1.5cm of h2] (h3);
        \vertex[right=1cm of h3] (h4);
        \vertex[right=1cm of h4] (c3);
       
        \vertex[right=1.5cm of c2] (f2);
        \vertex[right=1.2cm of f2] (f3);
        \vertex[above=0.9cm of f3] (g1);
        \vertex[below=0.9cm of f3] (t1);
        \vertex[right=1cm of g1] (c4) {\(Z\)};
        \vertex[right=1cm of t1] (c5) {\(Z\)};
        
       \diagram* {
        (a1) -- [gluon, thick] (a2);
        (b1) -- [gluon, thick] (b2);
        (a2) -- [fermion, thick, edge label=\(t\)] (b2) -- [fermion, thick, edge label'=\(t\)] (c2) -- [fermion, thick, edge label'=\(t\)] (a2);
        (c2) -- [scalar, thick,  edge label=\(h\)] (f2) -- [scalar,  edge label=\(h\), thick] (g1);
        (f2) -- [scalar,  edge label'=\(h\), thick] (t1) -- [boson,  edge label'=\(Z\), thick] (g1);
        (g1) -- [boson, thick] (c4);
        (t1) -- [boson, thick] (c5);
         };
         \filldraw[color=black, fill=black] (a2) circle (0.04);
         \filldraw[color=black, fill=black] (b2) circle (0.04);
        \filldraw[color=black, fill=black] (c2) circle (0.04);
         \filldraw[color=black, fill=black] (g1) circle (0.04);
          \filldraw[color=black, fill=black] (t1) circle (0.04);
        \node[square dot,fill=black,large] (d) at (f2){};
      \end{feynman}
\end{tikzpicture}
\end{subfigure}
\vspace{4mm}
\caption{\label{fig:diagrams} Representative Feynman diagrams that lead to a ${\cal O} (\lambda)$ correction to the process $gg \to h^\ast \to Z Z$. The black boxes denote insertions of the operator ${\cal O}_6$ introduced in~(\ref{eq:operatorsd6}). Consult the main text for further details.}
\end{figure}
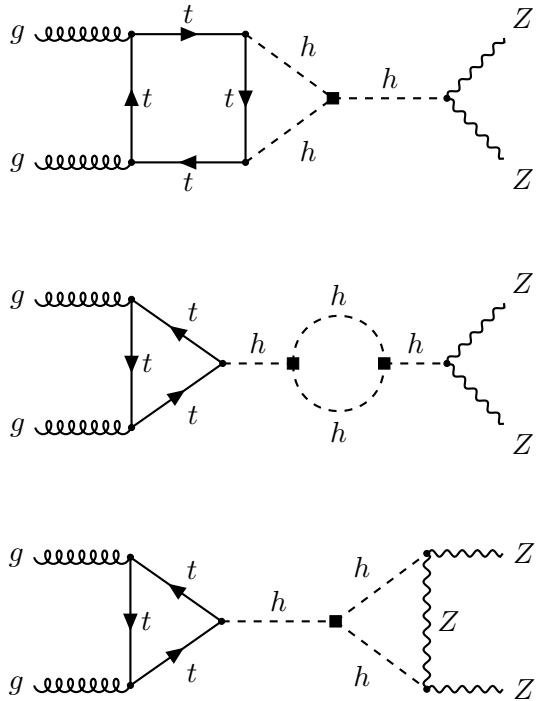

The  $\mathcal{O}(\lambda)$ corrections to ggF Higgs production receive contributions  from both two-loop topologies (see the upper Feynman diagram in Figure~\ref{fig:diagrams}) as well as from  the wave-function renormalisation of the Higgs boson field.  The relevant renormalised vertex that describes the process $g(p_1) + g (p_2) \to h(p_1+p_2)$ can be written as~\cite{Gorbahn:2019lwq}  
\beq \label{eq:Aggh}
\hat \Gamma_{ggh}^{\hspace{0.25mm} \mu \nu} \left ( p_1, p_2  \right ) = - \frac{\alpha_s \hspace{0.25mm}  \delta^{a_1 a_2}}{\pi v}   \hspace{0.5mm} \big ( \eta^{\mu \nu}\hspace{0.5mm}  p_1 \cdot p_2 - p_{1}^{\nu} \hspace{0.25mm}  p_{2}^{\mu} \big ) \hspace{0.5mm}   \Bigg[  \frac{ \delta Z_h}{2}  \hspace{0.5mm}  \mathcal{F}_1  + \frac{\lambda \bar{c}_6}{(4 \pi)^2} \hspace{0.5mm}  \mathcal{F}_2   \Bigg ]  \,,
\eeq
where $\alpha_s = g_s^2/(4 \pi)$ denotes the strong coupling constant, $a_1$ and $a_2$ are colour indices  and $\eta^{\mu \nu} = {\rm diag} \left ( 1, -1 , -1, -1 \right )$ is the Minkowski metric. In addition
\beq \label{eq:HWF}
\delta Z_h =  N_h \hspace{0.5mm} \bar c_6 \left (\bar{c}_6+2 \right ) \,, \qquad N_h = \frac{\lambda}{(4 \pi)^2} \hspace{-0.25mm} \left ( 9 - 2 \sqrt{3} \pi  \right ) \simeq -1.54 \cdot 10^{-3} \,,
\eeq
is the one-loop correction to the Higgs boson wave function associated to insertions of the operator ${\cal O}_6$~\cite{Gorbahn:2016uoy,Degrassi:2016wml} and 
\beq
{\cal F}_1 = \frac{m_t^2}{m_h^2} \hspace{0.25mm} \Big [ 2 - \left ( m_h^2 - 4 \hspace{0.125mm} m_t^2 \right ) C_0 \! \left ( \hat s , 0, 0, m_t^2, m_t^2, m_t^2 \right ) \Big  ] \,, 
\eeq
represents the one-loop  triangle diagram with internal  top quarks and $m_t$ is the top-quark mass. Here  $\hat s = 2  \hspace{0.25mm} p_1 \cdot p_2$  and the $C_0$ function denotes a three-point Passarino-Veltman scalar integral for which our definition follows the conventions used in the {\tt LoopTools} package~\cite{Hahn:1998yk}. The non-factorisable two-loop form factor~${\cal F}_2$ has been calculated analytically in~\cite{Degrassi:2016wml,Gorbahn:2019lwq} using the method of asymptotic expansions, which in this case is valid up to energies $\sqrt{\hat s} \simeq m_t$. To cover the full off-shell range of interest up to $\sqrt{\hat s } = 1 \, {\rm  TeV}$, we employ  the numerical results for the  non-factorisable  two-loop form factor~${\cal F}_2$ presented in~\cite{Bizon:2018syu,Borowka:2018pxx} in our numerical analysis performed in Section~\ref{sec:numan}. 

\subsection{Higgs propagator}

The Higgs propagator also receives corrections from insertions of the operator ${\cal O}_6$ (see the centre graph in Figure~\ref{fig:diagrams}). The resulting renormalised contribution to the self-energy of the Higgs takes the form
\begin{align}\label{eq:renprop}
\begin{split}
 \hat{\Sigma}(\hat s) =  \Sigma(\hat s) +  \left (\hat s - m_h^2 \right ) \delta Z_h -  \delta m_h^2  \ ,
\end{split}
\end{align}
where the one-loop corrections to the Higgs wave function has already been given in~(\ref{eq:HWF}) and the bare Higgs self-energy and the mass counterterm in the on-shell scheme are given by the following expressions
\begin{align} \label{eq:propct}
\begin{split}
\Sigma(\hat s) &= \frac{\lambda \bar{c}_6}{(4 \pi)^2}   \left (\bar{c}_6+2 \right ) 9 \hspace{0.125mm}  m_h^2  \hspace{0.125mm}B_0 \! \left (\hat s,m_h^2,m_h^2 \right )  \,, \\[2mm]
\delta m^2_h &=  \frac{\lambda \bar{c}_6}{(4 \pi)^2} \left (\bar{c}_6+2 \right)  9 \hspace{0.125mm} m_h^2  \hspace{0.125mm} B_0 \! \left (m_h^2,m_h^2,m_h^2 \right ) \,.
\end{split}
\end{align}
Here the $B_0$ functions are  two-point Passarino-Veltman scalar integrals defined as in~\cite{Hahn:1998yk}. 

\subsection{Higgs decay}

The full $\mathcal{O}(\lambda)$ correction to the Higgs decay $h \to ZZ$ receives a two-loop contribution (see the lower diagram in Figure~\ref{fig:diagrams}) as well as a counterterm contribution involving Higgs wave-function renormalisation. In the notation of~\cite{Bizon:2016wgr}, the relevant  renormalised vertex describing the $h (p_1+p_2) \to Z(p_1) Z(p_2)$ transition reads 
\begin{align} \label{eq:vertexZZ}
\begin{split}
\hat \Gamma^{\hspace{0.25mm} \mu \nu}_{hZZ} \left (p_1,p_2 \right )  &= \frac{2 m_Z^2}{v}  \Big [  \eta^{\mu \nu}   \mathcal{G}_1 + p_1^{\nu} p_2^{\mu} \hspace{.5mm} \mathcal{G}_2  \Big ]  \ ,
\end{split}
\end{align}
where $m_Z$ denotes the $Z$-boson mass. The $\mathcal{O}(\lambda)$ corrections to the  one-loop form factors $\mathcal{G}_1$ and $\mathcal{G}_2$ are given by 
\bea \label{eq:hzz_ff_PV}
\begin{split}
\mathcal{G}_1 &=  \frac{\delta Z_h}{2}  - \frac{\lambda \bar{c}_6}{(4 \pi)^2}  \hspace{0.25mm} \Bigg\{ 12 \hspace{0.25mm} \Big [ m_Z^2  \hspace{0.25mm} C_0 \! \left ( (p_1+p_2)^2, p_1^2, p_2^2, m_h^2, m_h^2, m_Z^2 \right )    \\[1mm]
& \hspace{1.9cm}   -  C_{00} \! \left ( (p_1+p_2)^2, p_1^2, p_2^2, m_h^2, m_h^2, m_Z^2 \right )  \Big ] +3  \hspace{0.125mm} B_0 \! \left ((p_1+p_2)^2, m_h^2, m_h^2 \right ) \!  \Bigg \} \ ,
\\[2mm] 
\mathcal{G}_2 &= \frac{\lambda \bar{c}_6}{(4 \pi)^2}   \hspace{0.5mm}   12 \hspace{0.25mm} \Big[C_1  \! \left ((p_1+p_2)^2, p_1^2, p_2^2, m_h^2, m_h^2, m_Z^2 \right )  + C_{11}  \! \left ((p_1+p_2)^2, p_1^2, p_2^2, m_h^2, m_h^2, m_Z^2 \right ) \hspace{2mm} \\[1mm]
& \hspace{1.9cm} + C_{12}  \! \left ((p_1+p_2)^2, p_1^2, p_2^2, m_h^2, m_h^2, m_Z^2 \right )   \Big] \,,
\end{split} 
\eea
and the tensor coefficients $C_1$, $C_{00}$ and $C_{11}$ of the three-point Passarino-Veltman integrals are defined as in the publications~\cite{Bizon:2016wgr,Hahn:1998yk}. 

\subsection{MC implementation}

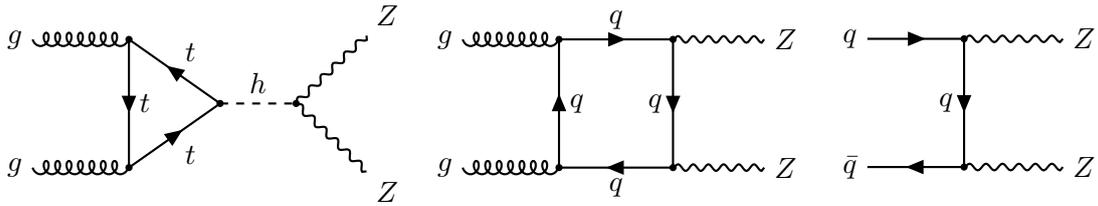
\begin{figure}[!t]
\centering
\begin{subfigure}[c]{0.375\linewidth}
\centering
 \begin{tikzpicture}
         \begin{feynman}
        \vertex (a1)  {\(g\)};  
        \vertex[right=1.5cm of a1] (a2);
        \vertex[below=1.7cm of a1] (b1)  {\(g\)};
        \vertex[right=1.5cm of b1] (b2);
        \vertex[below=0.85cm of a2] (c1);
        \vertex[right=1.2cm of c1] (c2);
        \vertex[right=1.1cm of c2] (h2);
        \vertex[right=1.5cm of h2] (h3);
        \vertex[right=1cm of h3] (h4);
        \vertex[right=1cm of h4] (c3);
       
        \vertex[right=1.0cm of c2] (f2);
        \vertex[right=1.2cm of f2] (f3);
        \vertex[above=0.9cm of f3] (g1) {\(Z\)};
        \vertex[below=0.9cm of f3] (t1) {\(Z\)};
              
       \diagram* {
        (a1) -- [gluon, thick] (a2);
        (b1) -- [gluon, thick] (b2);
        (a2) -- [fermion, thick, edge label=\(t\)] (b2) -- [fermion, thick, edge label'=\(t\)] (c2) -- [fermion, thick, edge label'=\(t\)] (a2);
        (c2) -- [scalar, thick,  edge label=\(h\)] (f2) -- [boson, thick] (g1);
        (f2) -- [boson, thick] (t1);
         };
         \filldraw[color=black, fill=black] (a2) circle (0.04);
         \filldraw[color=black, fill=black] (b2) circle (0.04);
         \filldraw[color=black, fill=black] (c2) circle (0.04);
         \filldraw[color=black, fill=black] (f2) circle (0.04);
          
      \end{feynman}
\end{tikzpicture}
\end{subfigure}
\begin{subfigure}[c]{0.3\linewidth}
\centering
 \begin{tikzpicture}
        \begin{feynman}
          \vertex (a1) {\(g\)};
          \vertex[below=1.7cm of a1] (a2){\(g\)};
          \vertex[right=1.5cm of a1] (a3);
          \vertex[right=1.5cm of a2] (a4);
          \vertex[right=1.5cm of a3] (a5);
          \vertex[right=1.2cm of a5] (a7) {\(Z\)};
          \vertex[right=1.5cm of a4] (a6);
          \vertex[right=1.2cm of a6] (a8) {\(Z\)};
          \diagram* {
             {[edges=gluon]
              (a1)--[thick](a3),
              (a2)--[thick](a4),
            },
            {[edges=fermion]
              (a3)--[thick, edge label=\(q\)](a5)--[thick, edge label'=\(q\)](a6)--[thick, edge label=\(q\)](a4)--[thick, edge label'=\(q\)](a3),
            },
            (a5) -- [boson, thick] (a7),
            (a6) -- [boson, thick] (a8),
          };
	 \filldraw[color=black, fill=black] (a3) circle (0.04);
	 \filldraw[color=black, fill=black] (a4) circle (0.04);
	 \filldraw[color=black, fill=black] (a5) circle (0.04);
	 \filldraw[color=black, fill=black] (a6) circle (0.04);
        \end{feynman}
      \end{tikzpicture}
\end{subfigure}
\quad 
\begin{subfigure}[c]{0.275\linewidth}
\centering
  \begin{tikzpicture}
        \begin{feynman}
          \vertex (a1) {\(q\)};
          \vertex[below=1.7cm of a1] (a2){\(\bar q\)};
          \vertex[right=1.5cm of a1] (a3);
          \vertex[right=1.5cm of a2] (a4);
          \vertex[right=1.3cm of a3] (a5) {\(Z\)};        
          \vertex[right=1.3cm of a4] (a6) {\(Z\)};
          \diagram* {
            {[edges=fermion]
              (a1)--[thick] (a3)--[thick, edge label'=\(q\)](a4)--[thick](a2),
            },
            (a3) -- [boson, thick] (a5),
            (a4) -- [boson, thick] (a6),
          };
	\filldraw[color=black, fill=black] (a3) circle (0.04);
	\filldraw[color=black, fill=black] (a4) circle (0.04);
        \end{feynman}
      \end{tikzpicture}
\end{subfigure}
\vspace{4mm}
\caption{SM contributions to $pp \to Z Z$. The left, centre and right  Feynman diagram represents a LO contribution to the $gg \to h^\ast \to Z Z$,  $gg \to Z Z$  and $q \bar q \to Z Z$ channel, respectively. In the case of the left graph there are also contributions involving bottom quarks. These diagrams are included in our analysis. See~the text for additional details.}
\label{fig:background}
\end{figure}

The three different types of $\mathcal{O}(\lambda)$ corrections affect not only the overall size of the $gg \to h^\ast \to Z Z  \to 4 \ell$ cross section, but also modify the shape of kinematic distributions such as the  four-lepton invariant mass $m_{4 \ell}$. In order to be able to predict $gg \to h^\ast \to Z Z  \to 4 \ell$  we have implemented the $\bar c_6$ corrections arising  from ggF Higgs production,  the Higgs propagator and the Higgs decay to $Z$ bosons into version 8.0 of {\tt MCFM}~\cite{Boughezal:2016wmq}. Our implementation includes all contributions up to ${\cal O} (\lambda^2)$  that arise from  squaring the full $gg \to h^\ast \to Z Z$ matrix element~(ME) which comprises both the BSM graphs depicted in Figure~\ref{fig:diagrams} as well as the LO SM  Feynman diagram shown on the left-hand side of Figure~\ref{fig:background}.  We also note that the contribution to the wave-function renormalisation constant~(\ref{eq:HWF}) coming from the propagator corrections at $\mathcal{O}(\lambda)$ exactly cancels against those of the vertices when combined to obtain  the full BSM contribution to the off-shell $gg \to h^\ast \to Z Z$ amplitude. This cancellation is expected since in the considered process the Higgs propagates on an internal line. It can be explicitly  seen by comparing the $\delta Z_h$ part of~(\ref{eq:renprop}), which contributes as $(- \delta Z_h  )$ times the LO SM amplitude since it is part of the second-order correction in the geometric series expansion of the propagator, with the $\delta Z_h$-dependent parts of the vertex contributions in~(\ref{eq:Aggh}) and~(\ref{eq:hzz_ff_PV}), which each gives a contribution proportional to $ \delta Z_h / 2$ times the LO SM amplitude. Notice that as a result of this cancellation the only ${\cal O} (\lambda)$ contributions quadratic in the Wilson coefficient~$\bar c_6$ arise from~(\ref{eq:propct}) with  the bare self-energy~$\Sigma (\hat s)$ being the only $\hat s$-dependent correction of this type. 

\section{\boldmath Numerical analysis}
\label{sec:numan}

In this section, we study the impact of the $\mathcal{O}(\lambda)$ corrections to the  $gg \to h^\ast \to Z Z  \to 4 \ell$  process   on the $m_{4 \ell}$ spectrum and a ME kinematic discriminant to be defined later. Following a brief discussion of the impact of QCD corrections, these results are then used to perform a sensitivity study of the two-dimensional constraints on the Wilson coefficients $\bar c_6$ and $\bar c_H$~---~see~(\ref{eq:lagrangian})~--- that can be obtained at LHC~Run~3 and the HL-LHC from measurements of $pp \to ZZ \to 4 \ell$.  The bounds from Higgs off-shell measurements are finally compared to the limits that are expected to arise from a combination of inclusive single-Higgs measurements at the end of LHC~Run~3 as well as the HL-LHC era. 

\subsection{Modifications of differential distributions}
\label{sec:modifications}

\begin{figure}[!t]
\begin{center}
\includegraphics[width=0.475\textwidth]{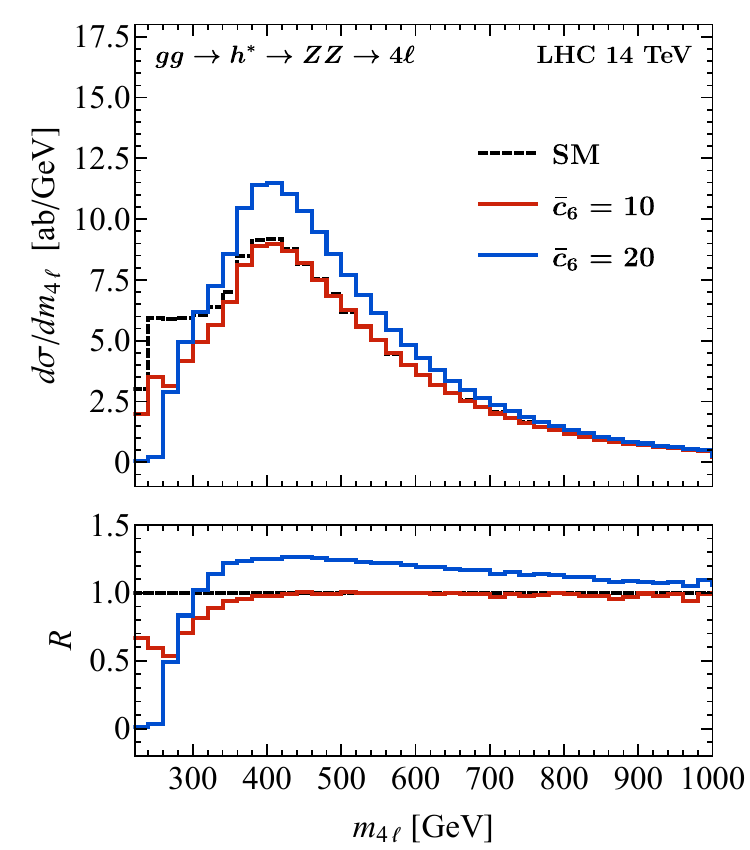} \quad 
\includegraphics[width=0.475\textwidth]{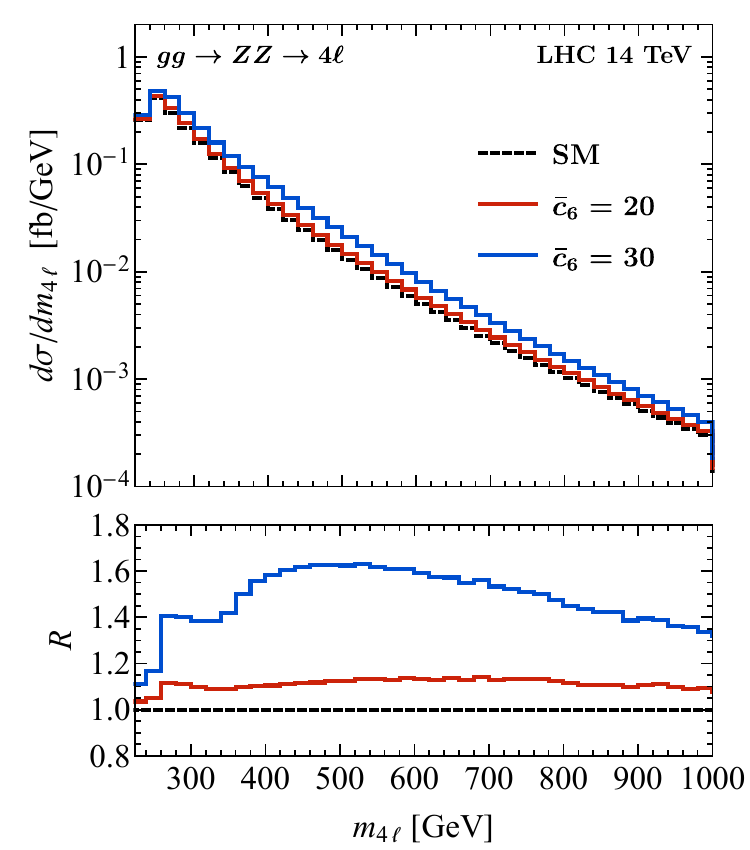}
\vspace{2mm} 
\caption{\label{fig:m4lplot} Left:  $m_{4 \ell}$~distributions for the Higgs signal in the~SM~(dashed black) and including the ${\cal O} (\lambda)$ corrections calculated in Section~\ref{sec:BSMcorrs} assuming $\bar{c}_6=10$~(solid red) and $\bar{c}_6=20$~(solid blue). Right:~$m_{4 \ell}$~spectra including the Higgs signal, the continuum background and their interference in the SM~(dashed black). The~results for the two BSM scenarios corresponding to $\bar{c}_6=20$~(solid red) and $\bar{c}_6=30$~(solid blue) are also displayed. All~distributions are obtained at LO in QCD assuming  $pp$ collisions at $\sqrt{s} = 14 \, {\rm TeV}$. The~lower panels show the ratios between the distributions and the corresponding SM predictions. Consult the main text for further details.}
\end{center}
\end{figure}

In Figure~\ref{fig:m4lplot} we show $m_{4 \ell}$ distributions for the Higgs channel alone~(left~panel) and for the Higgs channel, the gluon continuum background and their interference combined~(right panel) at LO in QCD. An example of a one-loop Feynman diagram that contributes to the SM $gg \to Z Z$ background is shown in the centre of~Figure~\ref{fig:background}.   We restrict ourselves to the off-shell region by considering a mass window of   $220 \, {\rm GeV} <m_{4 \ell}< 1000 \, {\rm GeV}$. The~leptons~($\ell = e, \mu$) are required to be  measured in the pseudorapidity range $|\eta_\ell| < 2.5$ and the lepton with the highest transverse momentum ($p_T$)  must satisfy $p_{T,\ell_1} > 20 \, {\rm GeV}$ while the second, third and fourth  lepton in $p_T$ order is required to obey  $p_{T,\ell_2} > 15 \, {\rm GeV}$, $p_{T,\ell_3} > 10 \, {\rm GeV}$ and $p_{T,\ell_4} > 6 \, {\rm GeV}$, respectively.  The lepton pair with the mass closest to the $Z$-boson mass is referred to as the leading dilepton pair and its invariant mass is required to be within $50 \, {\rm GeV} < m_{12} < 106 \, {\rm GeV}$, while the  subleading lepton pair must be in the range of $50 \, {\rm GeV} < m_{34} < 115 \, {\rm GeV}$. Similar cuts are employed in the ATLAS and CMS analyses~\cite{CMS:2014quz,ATLAS:2015cuo,ATL-PHYS-PUB-2015-024,CMS-PAS-FTR-18-011,ATLAS:2018jym,CMS:2019ekd,ATLAS:2019qet}. The  input parameters  used throughout our work are given by $G_F = 1/(\sqrt{2} \hspace{0.25mm} v^2) = 1.16639 \hspace{0.25mm}\cdot \hspace{0.25mm}10^{-5}\,  {\rm GeV}^{-2}$, $m_Z = 91.1876 \ \text{GeV}$, $m_h = 125 \,  {\rm GeV}$ and $m_t = 173 \, {\rm GeV}$.  The shown spectra assume $pp$ collisions at $\sqrt{s} = 14 \, {\rm TeV}$ and employ {\tt NNPDF40\_nlo\_as\_01180} parton distribution functions~(PDFs)~\cite{Ball:2021leu} with the renormalisation and factorisation scales $\mu_R$ and $\mu_F$ dynamically,~i.e.~for each event, set to $m_{4\ell}$. Our predictions include both the different-flavour (i.e.~$e^+ e^- \mu^+ \mu^-$) and the same-flavour (i.e.~$2e^+ 2e^-$ and $2\mu^+ 2 \mu^-$) decay channels of the two $Z$ bosons.

Two features of the distributions shown on the left-hand side of Figure~\ref{fig:m4lplot} deserve a further discussion. First, below the two-Higgs production threshold at $m_{4 \ell}=2 m_h$  the  BSM $gg \to h^\ast \to Z Z \to 4 \ell$ spectra are both visibly smaller that the SM prediction.   This feature can be understood by noting that for sufficiently large values of the Wilson coefficient $\bar c_6$ the terms~(\ref{eq:propct}) provide the dominant ${\cal O} (\lambda)$ corrections.  In~fact, the Higgs propagator corrections necessarily reduce the real part of $gg \to h^\ast \to Z Z \to 4 \ell$ amplitude for $m_{4\ell} < 2 m_h$, and this destructive interference can be so pronounced that the BSM contribution almost exactly cancels the SM contribution. Second, above the two-Higgs production threshold the bare Higgs self-energy $\Sigma (\hat s)$ develops an imaginary part because both internal Higgs lines in the propagator correction (see the  centre graph in Figure~\ref{fig:diagrams}) can go  on their mass shell. As a result the BSM $gg \to h^\ast \to Z Z \to 4 \ell$ spectra can be larger than  the SM prediction for $m_{4\ell} > 2 m_h$. Notice that the two aforementioned features are clearly visible in the case of the BSM distribution corresponding to $\bar c_6 = 20$, while for $\bar c_6 = 10$ the spectrum is not enhanced above the two-Higgs production threshold because the ${\cal O} (\lambda)$ corrections to Higgs production and decay (see the  upper and lower graph in~Figure~\ref{fig:diagrams}) are  numerically relevant in this case and tend to cancel the effect of the Higgs propagator. 

Our results for the  $gg  \to Z Z \to 4 \ell$ distributions including the Higgs signal, the continuum background and their interference are displayed in the right panel of Figure~\ref{fig:m4lplot}. In the vicinity of the two-Higgs production threshold $m_{4 \ell}=2 m_h$  one observes a plateau-like structure in both BSM spectra. This feature  arise from the combination of the modified Higgs signal and the interference of the BSM signal with the continuum SM background. This atypical shape change provides a genuine probe of loop corrections to the Higgs propagator involving light virtual particles. Such corrections arise in the case at hand  from the insertions of the operator~${\cal O}_6$ but  they can also appear in ultraviolet (UV) complete models of BSM physics (see for instance~\cite{Englert:2014aca,Englert:2014ffa,Goncalves:2017iub,Goncalves:2018pkt}). Both BSM spectra also show an enhancement in the $m_{4 \ell}$ tail.  Notice that the observed shape change is qualitatively different from  the  relative modifications that  arise from tree-level insertions of dimension-six SMEFT operators  which typically show a  roughly quadratic growth with~$m_{4 \ell}$~(see for example~\cite{Gainer:2014hha,Englert:2014ffa}). Similar statements apply to the case when the SM Higgs boson width is rescaled in such a way that the $pp \to h^\ast \to Z Z \to 4 \ell$ cross section close to the  Higgs  peak is unchanged~\cite{Kauer:2012hd,Caola:2013yja,Campbell:2013una}. See~Appendix~\ref{app:higgswidth} for a  detailed discussion. 

We have seen that the inclusion of the ${\cal O} (\lambda)$ corrections to the $pp \to h^\ast \to Z Z \to 4 \ell$ amplitude associated  to insertions of the SMEFT operator ${\cal O}_6$ lead to phenomenologically relevant kinematic features in the $m_{4\ell}$ distribution. The analysis sensitivity to the Higgs channel, especially in the off-peak region, has been shown to benefit considerably from the use of ME-based kinematic discriminants~(see for instance~\cite{Gao:2010qx,Bolognesi:2012mm,Anderson:2013afp,Campbell:2013una,CMS:2014quz,ATLAS:2015cuo,ATL-PHYS-PUB-2015-024,CMS-PAS-FTR-18-011,ATLAS:2018jym,CMS:2019ekd,ATLAS:2019qet}) to separate the $gg \to h^\ast \to Z Z \to 4 \ell$ signal from the main SM background coming from $ZZ$ production in $q \bar q$-annihilation.  A relevant  Feynman diagram that contributes to the  $q \bar q \to Z Z$ background   at LO in QCD is displayed on the right in~Figure~\ref{fig:background}. 
Being sensitive not only to $m_{4 \ell}$ but to another seven variables such as the invariant masses of the two opposite-sign lepton pairs (for details consult~\cite{Gao:2010qx,Bolognesi:2012mm,Anderson:2013afp}), the ME-based discriminants fully exploit the event kinematics.  In practice,  the ME-based discriminants are  often embedded in a multivariate discriminant based on a boosted decision tree~(BDT) algorithm, but as it turns out in the case of the four-lepton final state the sensitivity of the BDT analysis  improves only very little with respect to the ME-based discriminant alone (for example  in the case of the analysis~\cite{ATLAS:2015cuo} the improvement amounts to a mere 2\%). In the following, we restrict ourselves to an approach with only a  ME-based  discriminant, which we define as follows~\cite{ATLAS:2015cuo,ATL-PHYS-PUB-2015-024,ATLAS:2018jym,ATLAS:2019qet}
\begin{align}\label{eq:kindiscr}
D_S = \log_{10} \Bigg( \frac{P_h}{P_{gg} + c \cdot P_{q\bar{q}}} \Bigg) \, .
\end{align}
Here $P_h$ denotes the squared ME  for the $gg \to h^\ast \to Z Z \to 4 \ell$ process, $P_{gg}$ is the squared ME  for all  $gg$-initiated channels (including the Higgs channel, the continuum background and their interference) and $P_{q\bar{q}}$ is the squared ME  for the $q\bar{q} \to Z Z \to 4\ell$ process. Following the publications~\cite{ATLAS:2015cuo,ATL-PHYS-PUB-2015-024,ATLAS:2018jym,ATLAS:2019qet} the constant $c$ is set to 0.1 to balance the $q\bar{q}$- and $gg$-initiated contributions.
We add that in the SM more than 99\% of the $pp \to Z Z \to 4 \ell$  cross section fall into the range of $-4.5 < D_S < 0.5$~\cite{ATLAS:2015cuo}. The kinematic discriminant~(\ref{eq:kindiscr}) thus presents a null test for BSM models that lead to events with $D_S < -4.5$ or $D_S > 0.5$. 

\begin{figure}[!t]
\begin{center}
\includegraphics[width=0.475\textwidth]{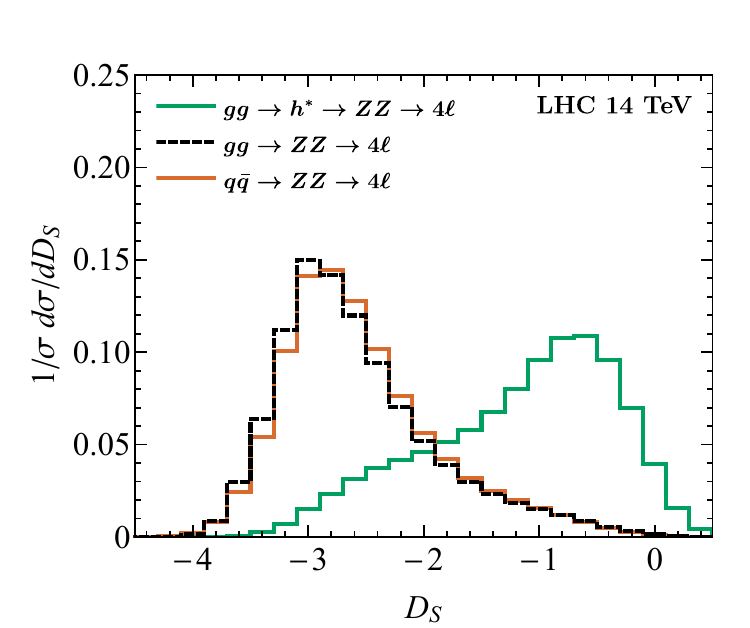} \quad 
\includegraphics[width=0.475\textwidth]{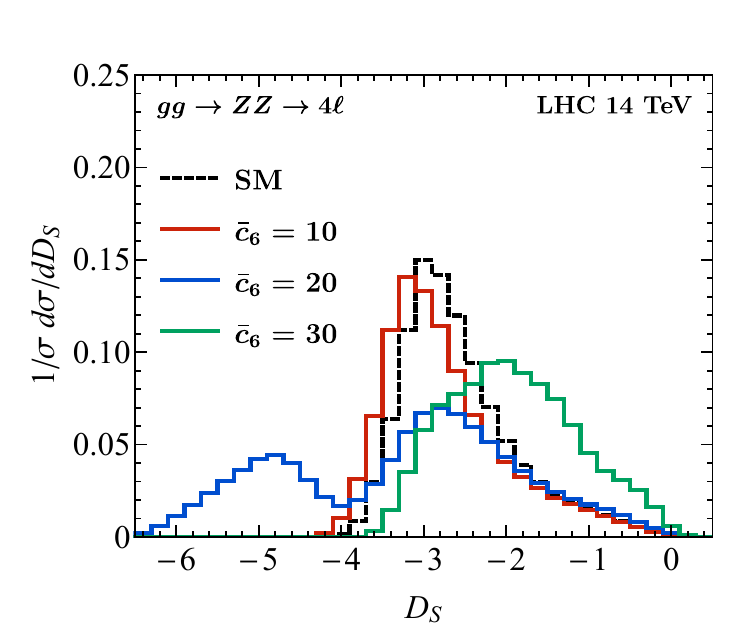}
\vspace{2mm} 
\caption{\label{fig:DSplot} Left:~Normalised $D_S$ distributions in the SM for the Higgs signal~(solid green), all~$gg$-initiated channels combined~(dashed black) and the $q\bar{q}$-initiated  background~(solid orange). Right:~Normalised $D_S$ distributions for the $gg$-initiated contributions in the SM~(dashed black) and for $\bar{c}_6=10$~(solid red), $\bar{c}_6=20$~(solid blue) and $\bar{c}_6=30$~(solid green). See the main text for further explanations.}
\end{center}
\end{figure}

To illustrate the discriminating power of the variable $D_S$ we show in the left panel of Figure~\ref{fig:DSplot} the normalised SM distributions at LO in QCD for the three contributions corresponding to the MEs that enter~(\ref{eq:kindiscr}). The discriminant $D_S$, which is calculated for every event in the simulation, is implemented in \texttt{MCFM} which uses the $gg$-initiated MEs provided in~\cite{Campbell:2013una}. One observes that the distribution corresponding to $q \bar q \to Z Z \to 4 \ell$ peaks at $D_S \simeq -3$, while the $g g \to h^\ast \to Z Z \to 4 \ell$ spectrum is shifted to higher $D_S$ featuring  a maximum at around $D_S \simeq -0.5$. An enhancement of the $g g \to h^\ast \to Z Z \to 4 \ell$ amplitude in the off-shell region will hence lead to a $D_S$ distribution of the full $pp \to Z Z \to 4 \ell$  process that is shifted to the right compared to the SM spectrum. For BSM scenarios that predict an enhancement in the tail of the $m_{4 \ell}$ distribution, one thus expects to find an excess of events for $D_S \gtrsim -1$. As discussed in Appendix~\ref{app:higgswidth}, this is precisely what happens if the Higgs boson couplings and its total  decay width are modified according to~(\ref{eq:rescale}). 

In the case of the ${\cal O} (\lambda)$ corrections to off-shell Higgs production resulting from the insertions of the SMEFT operator ${\cal O}_6$, the modifications of the normalised $D_S$ distributions are more intricate. For the full $gg \to Z Z \to 4 \ell$ contribution this is shown on the right-hand side of Figure~\ref{fig:DSplot}.  One observes that in the case of  $\bar c_6 = 10$ the BSM spectrum is shifted  to lower values compared to the SM result. This is a simple consequence of the fact that for $\bar c_6 = 10$ the $gg \to h^\ast \to Z Z \to 4 \ell$ amplitude and therefore $P_h$ in~(\ref{eq:kindiscr}) is reduced for $m_{4 \ell}$ values below the two-Higgs production threshold (cf.~the left panel in Figure~\ref{fig:m4lplot}). For $\bar c_6 = 20$ one instead sees that the $D_S$ distribution has two maxima: one at  $D_S \simeq -5$ and another one at $D_S \simeq -3$.  The peak at $D_S \simeq -5$ ($D_S \simeq -3$) is associated to kinematic configurations that lead to a reduction (an enhancement) of $P_h$ for $m_{4 \ell} < 2 m_h$ ($m_{4 \ell} > 2 m_h$) --- see again the left plot in Figure~\ref{fig:m4lplot}. Finally, for $\bar c_6 = 30$ the BSM contributions always enhance $P_h$ and hence the $D_S$ distribution is shifted to larger values. Notice that  the sharp cut-off of the SM distribution at $D_S \simeq -3.5$, makes~(\ref{eq:kindiscr})  more sensitive to kinematic configurations that reduce $P_h$ rather than enhance it. In fact, this feature turns out to be the key element that allows to set powerful constraints on BSM scenarios with $\bar c_6 \neq 0$ by means of the ME-based discriminant $D_S$. 

\subsection{Impact of QCD corrections}
\label{sec:QCD}

In the following we discuss the possible impact of higher-order QCD corrections to the $pp \to Z Z \to 4 \ell$ cross section differential in the  ME-based  kinematic distribution introduced in~(\ref{eq:kindiscr}). In~fact, as the ggF contribution to $pp \to Z Z \to 4\ell$ is loop-induced (see the centre graph in Figure~\ref{fig:background}), it enters the $ZZ$ production cross section at $\mathcal{O}(\alpha_s^2)$,~i.e.~at the next-to-next-to-leading order (NNLO) in QCD. State-of-the-art predictions for four-lepton production at the LHC, obtained at NNLO in QCD~\cite{Cascioli:2014yka,Grazzini:2015hta,Heinrich:2017bvg,Grazzini:2018owa}  and matched to parton shower~\cite{Alioli:2021wpn,Buonocore:2021fnj}, are reaching an impressive accuracy of $\mathcal{O}(2\%)$ for inclusive cross sections and $\mathcal{O}(5\%)$ in the case of differential distributions. 

While a precision phenomenological study of $ZZ$ production a la~\cite{Cascioli:2014yka,Grazzini:2015hta,Caola:2015psa,Heinrich:2017bvg,Grazzini:2018owa,Grazzini:2021iae,Alioli:2021wpn,Buonocore:2021fnj}  including the~${\cal O} (\lambda)$ corrections associated to the insertions of ${\cal O}_6$ is beyond the scope of this work,  we wish to assess at least approximately the impact of higher-order QCD corrections on our analysis, in particular the effects on the $D_S$ spectra. To this purpose we proceed in the following way. We first calculate for each production channel the so-called  $K$-factor defined as the ratio between the fiducial cross section at a given order in QCD and  the corresponding LO QCD prediction. In the case of the $gg$-initiated contribution we employ the results of the recent work~\cite{Buonocore:2021fnj}, where one of us reported NLO QCD corrections to the corresponding four-lepton invariant mass spectrum~(see Figure 2 of that article). One observes that the ratio between the NLO and LO ggF predictions is essentially flat in the region of interest for this work,~i.e.~for values of the invariant mass within $220 \, {\rm GeV} < m_{4\ell} <  1000 \, {\rm GeV}$. By~averaging over the ratio of the NLO and LO $m_{4 \ell}$ spectra within the aforementioned $m_{4 \ell}$ window we find $K^{\rm NLO}_{gg}  = 1.83$, which is in line with the previous works~\cite{Caola:2015psa,Grazzini:2018owa,Grazzini:2021iae}. In the case of the $q \bar q$-initiated contribution we utilise the LO and NNLO QCD results obtained in~\cite{Grazzini:2018owa} using {\tt MATRIX}~\cite{Grazzini:2017mhc}. The relevant $K$-factor again turns out to be basically flat in $m_{4\ell}$, with a central value of $K^{\rm NNLO}_{q\bar q}  = 1.55$, which is in accordance with~\cite{Cascioli:2014yka}. 

The aforementioned $K$-factors are listed in Table~\ref{tab:kfactors}, along with the scale uncertainties in each production channel at the relevant order in QCD. They are then used to obtain a QCD-improved prediction  for the $D_S$ distributions, in the following way:
\beq \label{eq:procedure}
\begin{split}
\left ( \frac{d \sigma ( pp \to Z Z \to 4 \ell)}{dD_S} \right )_{\rm improved} & = K^{\rm NLO}_{gg}  \left ( \frac{d \sigma ( gg \to Z Z \to 4 \ell)}{dD_S} \right )_{\rm LO} \\[2mm]  & \phantom{xx} +  K^{\rm NNLO}_{q \bar q}  \left ( \frac{d \sigma ( q \bar q \to Z Z \to 4 \ell)}{dD_S} \right )_{\rm LO} \,.
\end{split}
\eeq
Here the label LO indicates that both the cross sections as well as the ME-based discriminant $D_S$ are calculated at LO in QCD. Clearly,~(\ref{eq:procedure}) only captures part of the higher-order QCD corrections to the $d\sigma/dD_S$ spectra, namely those that are associated to the differential cross sections $d \sigma$, but ignores beyond-LO effects to $d D_S$. To improve upon this approximation one would need to extend the calculation of the ME-based discriminant~(\ref{eq:kindiscr})  to the NLO in QCD. While achieving NLO accuracy  is in principle possible for ME-based discriminants~\cite{Alwall:2010cq,Campbell:2012cz,Martini:2015fsa}, the actual calculations are in practice complicated by the fact that they require the use of modified jet algorithms to map resolved and unresolved parton configurations onto the proper MEs --- see also~\cite{Gritsan:2016hjl} for a discussion of this point. Since NLO and NNLO QCD corrections to the shapes of kinematic distributions in $pp \to Z Z \to 4 \ell$ are small and often indistinguishable when compared to the associated theoretical uncertainties, it is expected that the LO discriminant $D_S$ as defined in~(\ref{eq:kindiscr}) maintains its discriminating power beyond the well-defined~LO. This renders the procedure~(\ref{eq:procedure})  a pragmatic and simple approach to incorporate higher-order QCD effects. 

\begin{table}[t!]
\begin{center}
\begin{tabular}{| c | c | c | c | c |} 
\hline 
 & $\sigma_{gg}^{\rm LO}$  & $\sigma_{gg}^{\rm NLO}$ & $\sigma_{q\bar{q}}^{\rm LO}$ & $\sigma_{q\bar{q}}^{\rm NNLO}$  \\[1mm]
\hline 
$K_i^{\rm order} $ & 1 & 1.83  & 1 &  1.55\\[1mm]
 \hline 
 $\Delta_i^{\rm scale}$ & $^{+27.7\%}_{-20.5\%}$ & $^{+14.8\%}_{-13.4\%}$& $^{+5.5\%}_{-6.4\%}$ & $^{+1.1\%}_{-1.2\%}$ \\[1mm]
\hline 
\end{tabular}
\end{center}
\caption{\label{tab:kfactors} QCD $K$-factors, defined as $K_i^{\rm order} = \sigma_{i}^{\rm order}/\sigma_i^{\rm LO}$, for the channels $i=gg$, $q\bar{q}$ at different orders in QCD along with the associated relative scale uncertainties $\Delta_i^{\rm scale}$ for each channel and QCD order. The numbers for the $gg$-initiated channel are obtained utilising results presented in~\cite{Buonocore:2021fnj}, while the results for the $q\bar{q}$-initiated channel are taken from~\cite{Grazzini:2018owa}. Consult the main text for further details.}
\end{table}

\begin{figure}[!t]
\begin{center}
\includegraphics[width=0.475\textwidth]{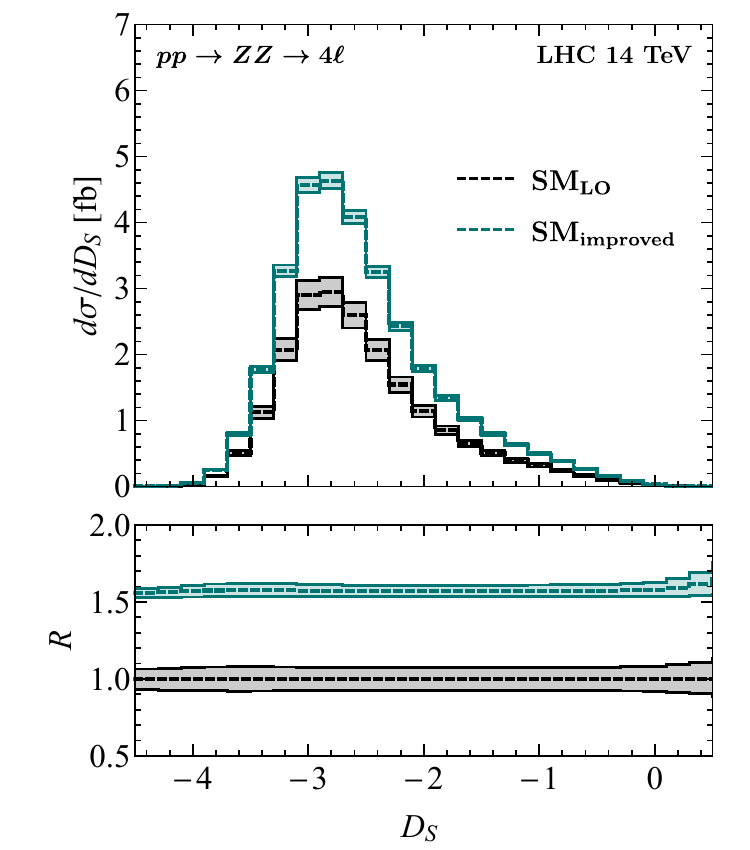} \quad 
\includegraphics[width=0.475\textwidth]{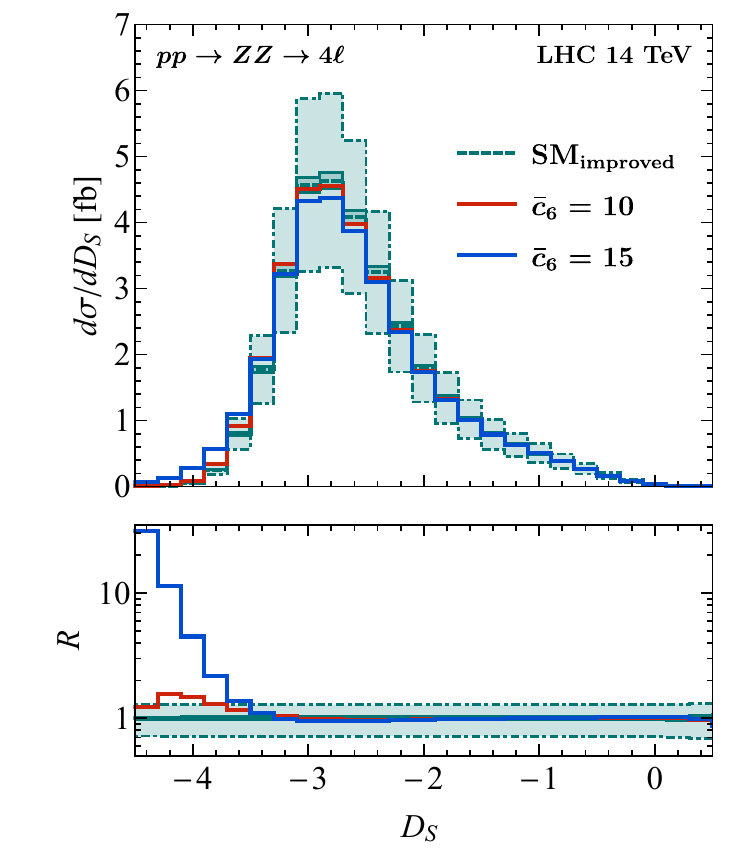}
\vspace{2mm} 
\caption{\label{fig:DSplotpp} Left: $D_S$~distributions for the $pp \to Z Z \to 4 \ell$  process in the SM obtained at~LO~(dashed black) and the QCD-improved prediction~(dashed green). The lower panel shows the ratio to the LO SM prediction and the uncertainty bands been obtained using the scale variations from Table~\ref{tab:kfactors}. Right:  $D_S$ spectra for  $pp \to Z Z \to 4 \ell$  production in the~SM~(dashed green) and for $\bar{c}_6=10$~(solid red) and  $\bar{c}_6=15$~(solid blue). The thin green band corresponds to the scale uncertainties of the QCD-improved SM predictions for~$D_S$, while the wide green band represents  half of the relative difference between the QCD-improved and the LO SM predictions for ME-based  discriminant. The lower panel shows the ratio of the BSM and SM predictions including the aforementioned uncertainty bands. All distributions  assume $pp$ collisions at $\sqrt{s} = 14 \, {\rm TeV}$ and the improved SM as well as the BSM predictions have  been obtained by means of~(\ref{eq:procedure}). Additional explanations can be found in the main~text. } 
\end{center}
\end{figure}

In the left plot of Figure~\ref{fig:DSplotpp} we compare the $D_S$ spectrum for $pp \to Z Z \to 4 \ell$ in the SM obtained at LO to the QCD-improved  $D_S$ spectrum~(\ref{eq:procedure}). One observes a close to flat $K$-factor of around~$1.6$ between the LO and the improved prediction of the $D_S$ spectrum for $pp \to Z Z \to 4 \ell$. The uncertainty bands represent the scale uncertainties obtained by applying the numerical values listed in Table~\ref{tab:kfactors} to the relevant the LO and QCD-improved spectra. We observe that the inclusion of higher-order QCD corrections reduces the scale uncertainties by a factor of about 3 from roughly~$(7 - 8)\%$ to about $(2 - 3)\%$, in line with recent precision SM phenomenological studies. However, the fact that the central value of the improved prediction lies well outside the LO uncertainty bands in all bins demonstrates that at LO, scale variations do not provide a reliable way to estimate the size of higher-order QCD effects. In fact, similar issues are known to occur also beyond~LO, for example in the case of inclusive ggF Higgs production~(see~\cite{Anastasiou:2015vya,Anastasiou:2016cez,Mistlberger:2018etf} for the corresponding state-of-the-art SM calculations) or even for four-lepton production~\cite{Buonocore:2021fnj},  where the loop-induced $gg$ contribution entering at NNLO is unaccounted for by the NLO scale uncertainties. With~this in mind and given  that the kinematic discriminant $D_S$ is only LO accurate, we wish to take a rather conservative approach and estimate the theoretical uncertainties to be half of the relative difference between the QCD-improved and the LO predictions, which corresponds to a relative uncertainty of about $30\%$.

On the right-hand side of Figure~\ref{fig:DSplotpp} we compare two BSM~$D_S$ distributions to the SM prediction, where the scale uncertainties quoted in Table~\ref{tab:kfactors} and the aforementioned conservative estimate are shown in the lower panel as the solid green and dot-dashed green bands around the SM spectrum, respectively.  The depicted  spectra have all been obtained using~(\ref{eq:procedure}). In accordance with the general discussion in Section~\ref{sec:modifications}, one observes that the~BSM spectra deviate most significantly from the SM distribution for $D_S \lesssim -3.5$, while for $D_S\gtrsim -3.5$ the deviations are generically small. In~fact, for the two choices of $\bar c_6$ shown in the figure the resulting modifications of the~$D_S$ spectrum for $D_S\gtrsim -3.5$ lie well within  our conservative theoretical uncertainty band. Furthermore, since the deviations for $D_S \lesssim -3.5$ are associated to the modification of the off-shell Higgs production cross section in the region $m_{4 \ell} < 2 m_h$, they will  lead  to a detectable change in events for a sufficiently large $|\bar c_6|$. For~instance, for $\bar c_6 = 10$ ($\bar c_6 = 15$) the shift in the  $pp \to Z Z \to 4 \ell$ cross section restricted to  $D_S < -3.5$ amounts to around $0.4 \, {\rm fb}$ ($1.2 \, {\rm fb}$)  compared to the SM. This corresponds to around 1200 (3700) additional $pp \to ZZ \to 4 \ell$ events at the HL-LHC. Notice finally that the relative modifications of the $D_S$ spectrum due to insertions of ${\cal O}_6$ are much larger than the shifts seen in the $m_{4 \ell}$ distribution~(cf.~the right panel in Figure~\ref{fig:m4lplot}). The ME-based discriminant~(\ref{eq:kindiscr}) therefore provides a significantly better sensitivity to BSM models with $\bar c_6 \neq 0$ than the $m_{4 \ell}$ spectrum. 

\subsection[Constraints on Wilson coefficients $\bar{c}_6$ and $\bar{c}_H$]{\boldmath Constraints on Wilson coefficients  $\bar{c}_6$ and $\bar{c}_H$}
\label{sec:c6chfit}

Below we determine the constraints on the Wilson coefficients $\bar{c}_6$ and $\bar{c}_H$ that  future LHC runs may be able to set. In the case of the constraints arising from off-shell Higgs production, we utilise the QCD-improved $D_S$ predictions obtained by~(\ref{eq:procedure}), assuming a detection efficiency of 99\% (95\%) for muons (electrons) that satisfy the event selections described at the beginning of Section~\ref{sec:modifications}. These efficiencies correspond to those reported  in the latest ATLAS analysis of off-shell Higgs production~\cite{ATLAS:2019qet}. The statistical uncertainties of  the computed $D_S$  distributions are determined per bin assuming Poisson statistics, i.e.~taking  the statistical error  to be $\sqrt{N_i}$ with $N_i$ the number of events in a given  bin $i$. The~largest systematic uncertainties in our analysis arise from the theoretical uncertainties on the $gg \to h^\ast \to ZZ \to 4 \ell$ signal process, the $gg \to ZZ \to 4 \ell$ and $q\bar q  \to ZZ \to 4 \ell$ background processes and the interference between the $gg$-initiated signal and background. For the theoretical uncertainties on the improved $D_S$ prediction~(\ref{eq:procedure}) we take the conservative estimate discussed in Section~\ref{sec:QCD}, in which we assume also PDF uncertainties at the level of $\pm 5\%$ are included. In our LHC~Run~3 analysis we will thus use a total theoretical uncertainty of $\pm 30\%$ when determining the bounds on  $\bar{c}_6$ and $\bar{c}_H$. Anticipating theoretical advances in the case of the HL-LHC we assume that the theoretical uncertainties due to scale variations and PDFs are reduced to $\pm 15\%$ which does no seem unrealistic. In fact, similar assumptions are made in the HL-LHC studies~\cite{ATL-PHYS-PUB-2015-024,CMS-PAS-FTR-18-011}.  Compared to the theoretical uncertainties the experimental uncertainties of systematic origin are close to negligible as they amount to ${\cal O} (2\%)$  (see for example~\cite{ATLAS:2014kct}). We will thus ignore the experimental systematics in what follows. 

The total number of events in bin $i$ depends on the Wilson coefficients $\bar c_6$ and $\bar c_H$ in the following way 
\beq \label{eq:Nic6cH}
N_i (\bar c_6, \bar c_H) = N_i (\bar c_6) - 2 \hspace{0.25mm} \bar c_H N_i (0) \,,
\eeq
where $N_i(\bar c_6)$  denotes the number of events which we calculate using {\tt MCFM} correcting them by QCD effects~(\ref{eq:procedure}) and lepton efficiencies.  Notice that $N_i(0)$ corresponds to the SM expectation of events.      The significance $Z_{i}$ is calculated as a Poisson ratio of likelihoods modified to incorporate systematic uncertainties on the background using the Asimov approximation~\cite{Cowan:2010js,CowanNotes}:
\beq \label{eq:Zi}
Z_{i} = \left \{ 2 \left [ \left (s_i + b_ i \right ) \ln \left [ \frac{\left (s_i + b_i \right) \left (b_i + \sigma_{b_i}^2 \right )}{b_i^2 + \left (s_i + b_i \right ) \sigma_{b_i}^2} \right ] - \frac{b_i^2}{\sigma_{b_i}^2}  \ln \left( 1+ \frac{s_i \hspace{0.25mm} \sigma_{b_i}^2}{b_i \hspace{0.25mm} (b_i + \sigma_{b_i}^2)} \right ) \right ]\ \right \}^{1/2} \,.
\eeq
Here $s_i$ ($b_i$) represents the expected number of signal (background) events in  bin $i$ and $\sigma_{b_i}$ denotes the standard deviation that characterises the systematic uncertainties of the associated  background. To set bounds on $\bar c_6$ and $\bar c_H$ we assume that the central values of a future measurements of the $D_S$ distribution will line up  with the SM predictions. We hence employ 
\beq \label{eq:sbsigma}
s_i = N_i (\bar c_6, \bar c_H) - N_i (0, 0) \,, \qquad b_i = N_i (0, 0) \,, \qquad \sigma_{b_i} =  \Delta_i N_i (0, 0)  \,,
\eeq 
where $\Delta_i$ denotes the relative  total systematic uncertainty in  bin $i$.  We will employ bin-independent systematic uncertainties of $\Delta_i = 0.3$ and $\Delta_i = 0.15$ at LHC~Run~3 and HL-LHC, respectively. The total significance $Z$ is obtained by adding the individual $Z_{i}$ values in quadrature. Parameter regions with a total significance of $Z > \sqrt{2} \hspace{0.5mm} {\rm erf}^{-1} \left ( {\rm CL} \right ) $ are said to be excluded at a given confidence level CL. Here ${\rm erf}^{-1} (z)$ denotes the inverse error function. In our numerical analysis, we include 29 bins of equal size of 0.2 that cover the range $-5.1 < D_S  <0.5$. 

\begin{figure}[!t]
\begin{center}
\includegraphics[width=0.475\textwidth]{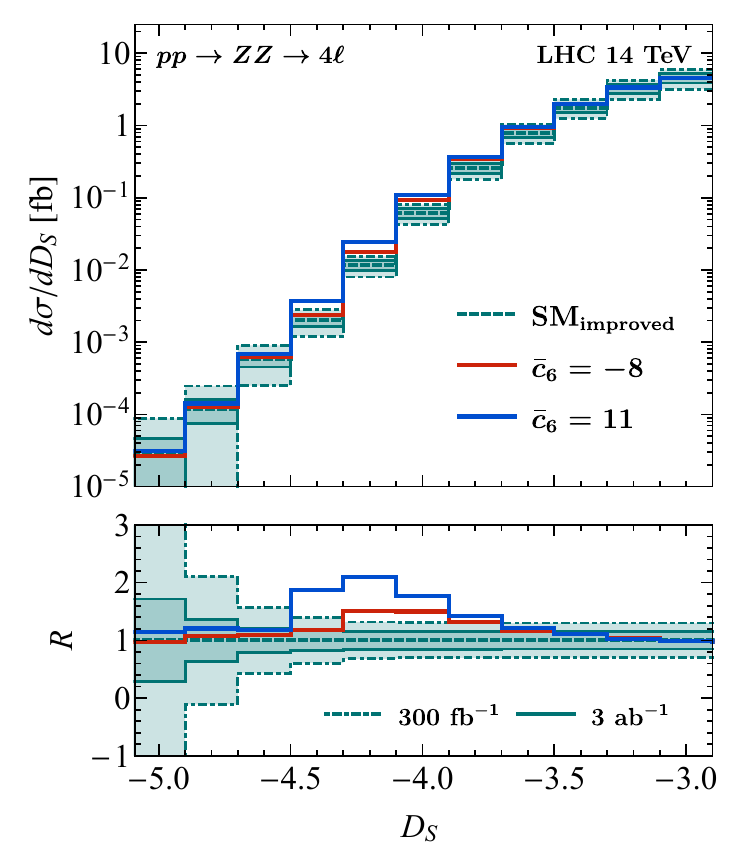} \quad 
\raisebox{5mm}{\includegraphics[width=0.475\textwidth]{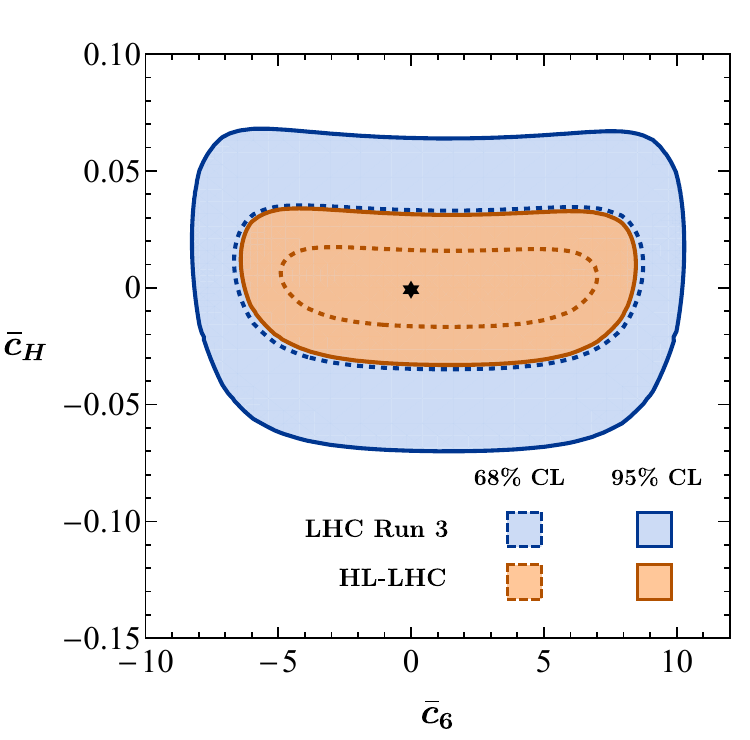}} 
\vspace{2mm} 
\caption{\label{fig:contours} Left: QCD-improved $D_S$ distributions for $pp \to Z Z \to 4 \ell$. The SM spectrum~(dashed green) and  BSM predictions for $\bar c_6 = -8$~(solid red) and  $\bar c_6 = 11$~(solid blue) are shown. The uncertainty bands around the SM spectrum indicate the total uncertainties expected at LHC~Run~3 and HL-LHC. Right: Projected 68\% and 95\%~CL constraints in the $\bar c_6 \hspace{0.25mm}$--$ \hspace{0.25mm} \bar c_H$ plane at LHC~Run~3~(dashed and solid blue) and HL-LHC~(dashed and solid red). The black star represents the SM prediction. For additional details see main text.} 
\end{center}
\end{figure}

Before deriving the projected bounds on the Wilson coefficient $\bar c_6$ from off-shell   Higgs production, we recall that the currently best LHC 95\%~CL limit reads~\cite{ATLAS-CONF-2019-049}
\beq \label{eq:LHCRun2}
\bar c_6 \in [-3.3,9.3] \,,\;\;  (\text{LHC Run 2}) \,. 
\eeq 
See also~\cite{Rossi:2020xzx,Degrassi:2021uik}. The bound~(\ref{eq:LHCRun2}) has been obtained from a combination of ten double-Higgs and single-Higgs production measurements performed by the ATLAS collaboration. Assuming $\bar c_H = 0$ and employing the fit strategy described above, we find the following 95\%~CL bounds from future hypothetical measurements of off-shell Higgs production:
\begin{align} \label{eq:c6boundsoff}
\bar{c}_6 \in [-8.2 , 10.2] \,, \;\; (\text{LHC Run 3}) \,,  \qquad  \bar{c}_6 \in [-6.3, 8.4] \,, \;\; (\text{HL-LHC}) \,.
\end{align}
The quoted limits for LHC~Run~3 (HL-LHC) correspond to  a full integrated luminosity $300 \, {\rm fb}^{-1}$ ($3 \, {\rm ab}^{-1}$) obtained at  $\sqrt{s} = 14 \, {\rm TeV}$. To~illustrate how the sensitivity of our fit to values of $|\bar c_6| = {\cal O}(10)$ arises,  we show on the left-hand side of Figure~\ref{fig:contours} the QCD-improved predictions for the $D_S$ spectrum within the SM and  two BSM models. The uncertainty bands around the SM prediction reflect the total uncertainties that are obtained by adding in quadrature the statistical and systematic uncertainties in each bin. From the figure it is evident that for both $\bar c_6 = -8$ and $\bar c_6 = 11$, the $D_S$ spectrum in enhanced over the SM background within the range $-4.5 \lesssim D_S \lesssim -3.5$. In this range the total uncertainties are largely dominated by the theory uncertainties with the statistically errors playing only a minor role. This feature allows to set  limits like~(\ref{eq:c6boundsoff})  that are competitive with~(\ref{eq:LHCRun2}) using solely off-shell Higgs production in the ggF channel. 

On the right in Figure~\ref{fig:contours} we furthermore show the  68\%~CL~(dashed) and 95\%~CL~(solid) constraints in the $\bar c_6 \hspace{0.25mm}$--$ \hspace{0.25mm} \bar c_H$ plane that derive from our fit employing LHC~Run~3 (blue) and HL-LHC (red) data. One observes that the bounds on $\bar c_H$ are essentially independent of the precise value of~$\bar c_6$ if the latter Wilson coefficient is sufficiently small. In the case of~$\bar c_6 = 0$, we obtain at 95\%~CL for instance 
\begin{align} \label{eq:cHboundsoff}
\bar{c}_H \in [-7.0,6.4] \cdot 10^{-2} \,, \;\; (\text{LHC Run 3}) \,,  \qquad  \bar{c}_H \in [-3.3, 3.1]  \cdot 10^{-2}  \,, \;\; (\text{HL-LHC}) \,.
\end{align}
We emphasise that non-zero values of  $\bar c_H$  do not change the shape of the $D_S$ spectrum  but only its normalisation~$\big($cf.~(\ref{eq:Nic6cH})$\big)$. This feature explains the approximate $\bar c_6\hspace{0.25mm}$-independence of the exclusion contours in the $\bar c_6 \hspace{0.25mm}$--$ \hspace{0.25mm} \bar c_H$ plane for small~$\bar c_6$. Notice finally that the bounds on~$\bar c_H$ are significantly more stringent than those on~$\bar c_6$. This is expected because the SMEFT operator~${\cal O}_H$ changes the prediction for $pp \to h^\ast \to ZZ \to 4 \ell$ already at LO  (i.e.~one loop) while  the corrections due to ${\cal O}_6$ start at NLO (i.e.~two loops). 

\subsection{Comparison to bounds from inclusive single-Higgs production}
\label{sec:onshcomp}

To further demonstrate the benefits of off-shell Higgs production in setting bounds on the Higgs trilinear coupling, we compare the results obtained in the previous section to the projected constraints one expects to obtain from inclusive single-Higgs measurements at future LHC runs. In the inclusive case the ${\cal O} (\lambda)$ corrections to the various Higgs production and decay channels can be written in terms of $\bar c_6$ and $\bar c_H$ as  
\begin{align} \label{eq:dsigmadBR}
\begin{split}
\delta \sigma_i  (\bar{c}_6,  \bar{c}_H)  =  \bar{c}_6 \Big( N_h \big( \bar{c}_6 + 2 \big) + C_1^{\sigma_i}    \Big)  - \bar{c}_H \,, \qquad 
\delta \text{BR}_f  (\bar{c}_6) = \frac{  \bar{c}_6 \left ( C_1^{\Gamma_f} - C_1^{\Gamma_{h}} \right )  }{1+\bar{c}_6 \hspace{0.25mm} C_1^{\Gamma_{h}} } \,,
\end{split}
\end{align}
where $N_h$ has been defined already in~(\ref{eq:HWF}) and $C_1^{\Gamma_{h}} = 0.23  \hspace{.25mm} \cdot \hspace{.25mm} 10^{-2}$~\cite{Degrassi:2016wml,Bizon:2016wgr,Gorbahn:2019lwq}. Notice that the Wilson coefficient $\bar c_H$ leads to a universal correction to all Higgs decay channels. Therefore it leaves the Higgs branching ratios unchanged. The calculations needed to obtain the process-dependent factors $C_1^{\sigma_i}$ and  $C_1^{\Gamma_f}$ have been performed in~\cite{Gorbahn:2016uoy,Degrassi:2016wml,Bizon:2016wgr,Maltoni:2017ims,Gorbahn:2019lwq,Degrassi:2019yix}. In our numerical analysis of the inclusive single-Higgs observables we include ggF, $Wh$, $Zh$, vector-boson fusion~(VBF) and $t \bar th$ production and consider the Higgs-boson branching ratios to pairs of photons ($\gamma \gamma$), EW bosons ($W^+W^-, ZZ$), bottom quarks  ($b \bar b$) and tau leptons ($\tau^+ \tau^-$). The~associated $C_1^{\sigma_i}$ and  $C_1^{\Gamma_f}$ coefficients are collected in~Table~\ref{tab:C1}.

In terms of~(\ref{eq:dsigmadBR}), keeping only terms linear in $\lambda$, the Higgs signal strengths for production in channel $i$ and decay in channel $f$ can be written   as
\begin{align} \label{eq:signalstrength}
\mu_i^f (\bar{c}_6,  \bar{c}_H)  = 1 + \delta \sigma_i (\bar{c}_6,  \bar{c}_H) + \delta \text{BR}_f (\bar{c}_6) \,, 
\end{align}
which we use to build the following $\chi^2$ function:
\begin{align} \label{eq:chi2}
\chi^2(\bar{c}_6, \bar{c}_H) = \sum_{i,f}\frac{\big (\mu_i^f(\bar{c}_6,  \bar{c}_H)-1 \big )^2}{\big (\Delta^f_i \big )^2} \,.
\end{align}
Here we have assumed that the central values of the future measurements of the Higgs signal strengths will coincide with the corresponding SM predictions. The variables $\Delta^f_i$ encode the relative total uncertainties obtained by combining the theoretical and statistical uncertainties associated to $\mu_i^f$. We collect the values of the  $\Delta^f_i$ used in our LHC~Run~3 and HL-LHC analyses in Table~\ref{tab:Deltaif}. Notice that the LHC~Run~3 numbers are obtained by combing the current theoretical and the statistical uncertainties in quadrature, while the HL-LHC numbers  assume that all theory uncertainties are halved with respect to our current understanding of  the relevant signals and backgrounds. The latter assumption corresponds to the scenario  S2 in the ATLAS paper~\cite{ATL-PHYS-PUB-2018-054}. The allowed CL regions are then obtained  by minimising~(\ref{eq:chi2}) and determining the solutions to $\Delta \chi^2  = \chi^2 -  \chi^2_{\rm min} < 2 \hspace{0.25mm} Q^{-1} (1/2, 1- {\rm CL})$ with $Q^{-1} (a, z)$ the regularised incomplete gamma function.

\begin{table}[t!]
\begin{center}
\begin{tabular}{| c | c | c | c | c | c |} 
\hline 
&ggF& $Wh$ & $Zh$ & VBF & $t\bar{t}h$ \\
\hline 
$C_1^{\sigma_i}$ & $0.66 \cdot 10^{-2}$ & $1.03 \cdot 10^{-2}$ & $1.18 \cdot 10^{-2}$ & $0.64 \cdot 10^{-2}$ & $3.47 \cdot 10^{-2}$ \\
\hline \hline 
& $\gamma \gamma$ & $W^+W^-$ & $ZZ$ & $b \bar b$ & $\tau^+ \tau^-$ \\
\hline 
$C_1^{\Gamma_f}$ & $0.49 \cdot 10^{-2}$ & $0.73 \cdot 10^{-2}$ & $0.83 \cdot 10^{-2}$ &  $0.67 \cdot 10^{-5}$ & $0.33 \cdot 10^{-5}$  \\
\hline 
\end{tabular}
\end{center}
\caption{\label{tab:C1} Values of the process-dependent coefficients $C_1^{\sigma_i}$ and  $C_1^{\Gamma_f}$. The numbers are directly taken or obtained from~\cite{Degrassi:2016wml,Bizon:2016wgr,Gorbahn:2019lwq} and the coefficients  $C_1^{\sigma_i}$  correspond to $pp$ collisions at  $\sqrt{s} = 14 \, {\rm TeV}$.}
\end{table}

\begin{table}[t!]
\begin{center}
\begin{tabular}{|c | c | c  |} 
\hline 
production, decay & LHC~Run~3 & HL-LHC  \\
\hline \hline 
ggF, $h \to \gamma\gamma$ & 0.13 &  0.036 \\
\hline 
ggF, $h \to W^+ W^-$ & 0.13 &  0.044 \\
\hline 
ggF, $h \to ZZ$ & 0.12 &  0.039 \\
\hline 
$Wh$, $h \to \gamma \gamma$ & 0.48 &  0.138 \\
\hline 
$Wh$, $h \to b \bar b$ & 0.57 &  0.100 \\
\hline 
$Zh$, $h \to \gamma \gamma$ & 0.85 & 0.157 \\
\hline 
$Zh$, $h \to b \bar b$ & 0.29 &  0.052 \\
\hline 
$Vh$, $h \to ZZ$ & 0.35 & 0.182 \\
\hline 
VBF, $h \to \gamma \gamma$ &  0.47 & 0.089 \\
\hline 
VBF, $h \to W^+ W^-$ & 0.21 & 0.066 \\
\hline 
VBF, $h \to Z Z$ & 0.36 & 0.118 \\
\hline 
VBF, $h \to \tau^+ \tau^-$ & 0.21 & 0.078 \\
\hline 
$t \bar t h$, $h \to \gamma \gamma$ & 0.38 & 0.074 \\
\hline 
$t \bar t h$, $h \to ZZ$ & 0.49 & 0.193 \\
\hline 
\end{tabular}
\end{center}
\caption{\label{tab:Deltaif} Relative total uncertainties  $\Delta^f_i$  on the Higgs signal strengths defined in~(\ref{eq:signalstrength}). The quoted  LHC~Run~3  and HL-LHC numbers are taken from~\cite{ATL-PHYS-PUB-2014-016} and~\cite{ATL-PHYS-PUB-2018-054}, respectively. Further information can be found in the main text. }
\end{table}

\begin{figure}[!t]
\begin{center}
\includegraphics[width=0.465\textwidth]{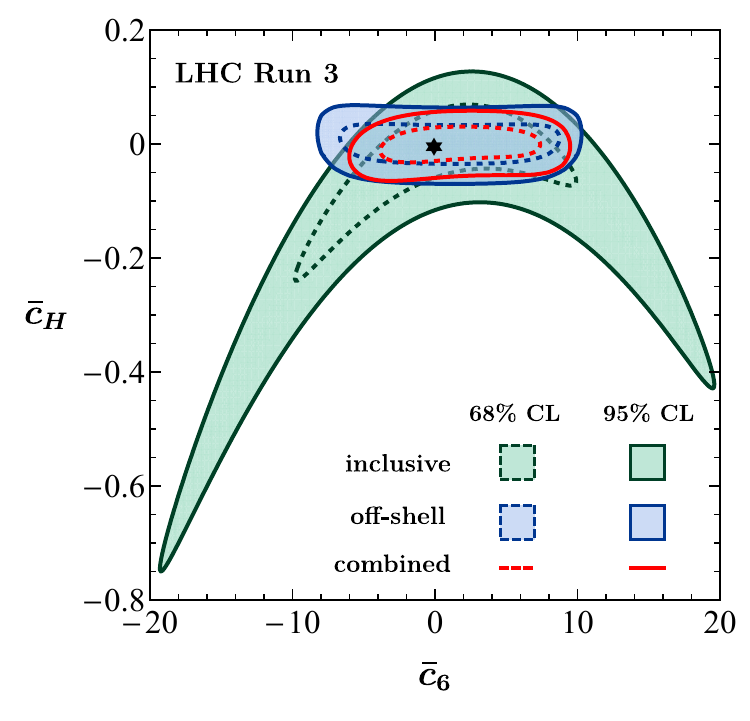} 
\quad
\includegraphics[width=0.475\textwidth]{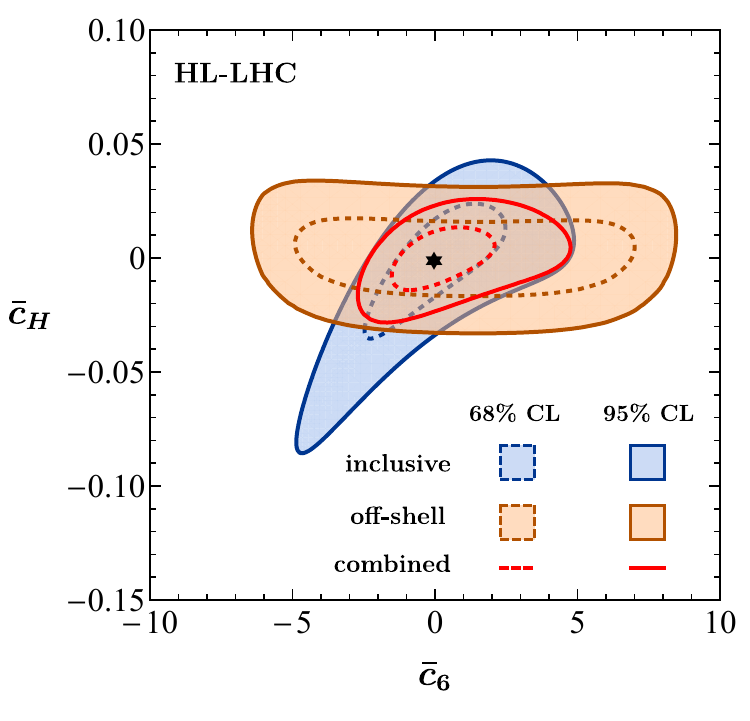}
\vspace{2mm} 
\caption{\label{fig:comparison_incl}  Projected 68\% and 95\%~CL constraints in the $\bar c_6 \hspace{0.25mm}$--$ \hspace{0.25mm} \bar c_H$ plane for the LHC~Run~3~(left) and the HL-LHC~(right) assuming integrated luminosities of 300~fb$^{-1}$ and 3~ab$^{-1}$, respectively, and $pp$ collisions at $\sqrt{s} = 14 \, {\rm TeV}$. The constraints from inclusive single-Higgs probes (left: green regions, right: blue regions) are compared to the off-shell Higgs constraints (left: blue regions, right: orange regions). The combinations of all constraints are also shown as red contours. The black stars represent the SM point. See main text for further explanations.} 
\end{center}
\end{figure}

In the left (right) panel of Figure~\ref{fig:comparison_incl} we show the projected 68\% and 95\%~CL constraints in the $\bar c_6 \hspace{0.25mm}$--$ \hspace{0.25mm} \bar c_H$ plane for the LHC~Run~3 (HL-LHC). One observes that the LHC~Run~3 fit to the~15 inclusive single-Higgs observable listed in~Table~\ref{tab:Deltaif} shows a pronounced flat direction in the $\bar c_6 \hspace{0.25mm}$--$ \hspace{0.25mm} \bar c_H$ plane. To understand this feature one first has to realise that only the process-independent coefficients $C_1^{\sigma_i}$ and $C_1^{\Gamma_f}$ are able to break flat directions in the inclusive fit. The relatively large coefficient $C_1^{\sigma_{t \bar t h}}$ (cf.~Table~\ref{tab:C1}) plays a particularly important role in this respect. Given the large total uncertainties of  $t \bar t h, h \to \gamma \gamma, ZZ$ at LHC~Run~3, the constraining power of the $t \bar t h$ channels  and thus the impact of $C_1^{\sigma_{t \bar t h}}$ is however limited. As~a result, the inclusive LHC~Run~3 exclusions are mainly determined by the ggF channels that  have a flat  direction for  $\bar c_6$ and $\bar c_H$ satisfying $\mu_{\rm ggF}^f \simeq \delta \sigma_{\rm ggF} (\bar c_6, \bar c_H ) \simeq 0$. The situation is visibly improved in the inclusive HL-LHC fit, mostly because the total uncertainties of the $t \bar t h$ channels are expected to be significantly reduced.  From both panels in Figure~\ref{fig:comparison_incl} it is however also evident that the flat direction in the inclusive fit is strongly broken by the  constraints arising from off-shell Higgs production. 

From the above it should be clear that inclusive single-Higgs and off-shell Higgs measurements should therefore be combined if one wants to exploit the full potential of the LHC in constraining the trilinear Higgs coupling through indirect probes. Performing such a combined analysis, we find for $\bar c_H =0$ the  following 95\%~CL limits 
\begin{align} \label{eq:c6boundsall}
\bar{c}_6 \in [-5.8 , 9.5] \,, \;\; (\text{LHC Run 3}) \,,  \qquad  \bar{c}_6 \in [-2.3, 4.6] \,, \;\; (\text{HL-LHC}) \,,
\end{align}
while for $\bar c_6 =0$ we obtain 
\begin{align} \label{eq:cHboundsall}
\bar{c}_H \in [-6.0, 5.6] \cdot 10^{-2} \,, \;\; (\text{LHC Run 3}) \,,  \qquad  \bar{c}_H \in [-2.3, 2.3]  \cdot 10^{-2}  \,, \;\; (\text{HL-LHC}) \,.
\end{align}
We add that the bounds~(\ref{eq:c6boundsall}) and~(\ref{eq:cHboundsall}) depend in a non-negligible way on the  assumed total uncertainties. In this respect one should remember that in the case of the constraints arising from off-shell Higgs production in ggF production we have assumed total systematic uncertainties of $\pm 30\%$ and $\pm 15\%$ in our LHC~Run~3 and HL-LHC fit, respectively. We believe that these are conservative uncertainties --- results for two additional  more aggressive assumptions about the systematic uncertainties entering the HL-LHC off-shell Higgs  analysis can be found in Appendix~\ref{app:HLLHCprojections}.  In fact, given the steady progress in the calculation of massive higher-loop corrections to $pp \to ZZ \to 4 \ell$ (see~\cite{Agarwal:2020dye,Bronnum-Hansen:2021olh} for the latest theoretical developments) and in view of the fact that it is theoretically known of how to achieve NLO accuracy for  ME-based discriminants~\cite{Alwall:2010cq,Campbell:2012cz,Martini:2015fsa}, it should be possible to put our naive estimates of theoretical uncertainties on more solid grounds. As can be seen from the left panel in Figure~\ref{fig:contours}, any improvement of our theoretical understanding of the $D_S$ distribution for  $D_S \lesssim -3.5$ will notably increase the sensitive of off-shell Higgs measurements to modifications of the trilinear Higgs coupling. 

\section{\boldmath Conclusions and outlook}
\label{sec:concl}

In this article, we have studied the constraints on the trilinear Higgs coupling that originate from off-shell Higgs production in $pp$ collisions at future LHC runs. To keep the discussion model-independent we have worked in the context of the SMEFT in which  the renormalisable SM interactions are augmented by the dimension-six operators ${\cal O}_6$ and ${\cal O}_H$ $\big($cf.~(\ref{eq:operatorsd6})$\big)$. Our computation of the $gg \to h^\ast \to Z Z \to 4 \ell$ process includes   two-loop corrections to ggF Higgs production and  one-loop corrections to the Higgs propagator as well as the decay $h^\ast \to ZZ$. The resulting scattering amplitudes have been implemented into {\tt MCFM} where they can be combined with the SM MEs for  $gg \to h^\ast \to ZZ \to 4 \ell$, $gg \to ZZ \to 4 \ell$ and $q \bar q\to ZZ \to 4 \ell$  to obtain differential distributions for the full $pp \to ZZ \to 4 \ell$ process, including the corrections due to insertions of the SMEFT operator ${\cal O}_6$. 

Using our MC implementation, we have then studied the shape differences in the four-lepton invariant mass $m_{4 \ell}$  distribution and the ME-based kinematic discriminant $D_S$ defined in~(\ref{eq:kindiscr}) that are induced by modifications of the  trilinear Higgs coupling. We found that the  inclusion of BSM effects leads to phenomenologically relevant kinematic features in both spectra.  In fact, the discriminant $D_S$ turns out to provide particularly powerful constraints on the Wilson coefficient~$\bar c_6$ of the pure-Higgs operator ${\cal O}_6$. The stringent constraints on~$\bar c_6$ arise because BSM scenarios  with $\bar c_6 \neq 0$    can lead to $D_S < -4.5$, which provides a null test since 99\% of the $pp \to ZZ \to 4 \ell$ events in the SM fall into the range $-4.5 < D_S < 0.5$. We~have also assessed the  possible impact of higher-order QCD correction to~$D_S$, arguing that~(\ref{eq:kindiscr}) maintains its discriminating power beyond the well-defined LO.

To  demonstrate the benefits of off-shell Higgs production in setting bounds on the Higgs trilinear coupling, we have determined the constraints on the Wilson coefficients $\bar c_6$ and $\bar c_H$ that the LHC with $300 \, {\rm fb}^{-1}$ and $3 \, {\rm ab}^{-1}$ of integrated luminosity at $\sqrt{s} = 14 \, {\rm TeV}$ may be able to set. We have then compared the obtained LHC Run~3 and HL-LHC bounds to the projected constraints that a combination of inclusive single-Higgs  measurements is expected to provide. Our analysis shows that ggF off-shell Higgs production  allow to put constraints on the trilinear Higgs coupling that are not only competitive with but also complementary to the exclusion limits obtained from inclusive single-Higgs production. Specifically, we found that future  studies of  the $D_S$ distribution in $pp \to ZZ \to 4 \ell$ production should help to remove flat directions in the $\bar c_6 \hspace{0.25mm}$--$ \hspace{0.25mm} \bar c_H$ plane that remain unresolved in fits that incorporate only inclusive single-Higgs measurements.  By combining all single-Higgs boson measurement we find that at the LHC~Run~3~(HL-LHC) it might be possible to constrain modifications of the trilinear Higgs coupling as parameterised by~(\ref{eq:c3})  to the 95\%~CL range $c_3  \in [-4.0, 6.1]$ ($c_3  \in [-1.7, 5.7]$). Additional HL-LHC projections that employ  two different assumptions about the systematic uncertainties entering our off-shell Higgs  analysis can be found in~Appendix~\ref{app:HLLHCprojections}. 

The studies performed in this article can be extended in several ways. First, to strengthen the constraints on the trilinear Higgs coupling derived in our work, one should also include in the projections measurements of double-Higgs production as well as EW precision observables. See~\cite{ATLAS-CONF-2019-049,Rossi:2020xzx,Degrassi:2021uik} for such global analyses based on LHC~Run~2 data. Second, the ME-based discriminant $D_S$ might also be a powerful tool to constrain BSM effects in $pp \to ZZ \to 4 \ell$ that arise in   the context of Higgs portal models~\cite{Englert:2014ffa,Goncalves:2017iub,Goncalves:2018pkt}. Given the limited direct LHC reach in such models~\cite{Ruhdorfer:2019utl,Haisch:2021ugv}, investigating the indirect sensitivity that off-shell Higgs production can provide seems like a worthwhile and timely exercise. Third, since any improvement of our theoretical understanding of the $D_S$ distribution~(\ref{eq:kindiscr}) will have a tangible impact on the sensitivity of off-shell Higgs measurements to modifications of the trilinear Higgs coupling, we believe that theoretical activity in this direction is important. Since many ingredients (if not all) are already available in the literature on $pp \to ZZ \to 4 \ell$~production~\cite{Cascioli:2014yka,Grazzini:2015hta,Caola:2015psa,Heinrich:2017bvg,Grazzini:2018owa,Grazzini:2021iae,Alioli:2021wpn,Buonocore:2021fnj,Alwall:2010cq,Campbell:2012cz,Martini:2015fsa,Agarwal:2020dye,Bronnum-Hansen:2021olh} achieving such a goal is certainly possible. 

\acknowledgments
We thank Darren Scott for useful discussions. The Feynman diagrams shown in this work were drawn with {\tt TikZ-Feynman}~\cite{Ellis:2016jkw}.

\appendix

\section{Higgs width effects}
\label{app:higgswidth}

In this appendix we illustrates how rescalings of the form 
\beq \label{eq:rescale}
g_{h XX}^{\rm SM} \to \xi^{1/4} \hspace{0.5mm} g_{h XX}^{\rm SM}  \,, \qquad
\Gamma_h^{\rm SM}\to \xi \hspace{0.5mm}   \Gamma_h^{\rm SM} \,, 
\eeq
with $g_{hXX}^{\rm SM}$ and $\Gamma_h^{\rm SM}$ denoting the couplings and total decay width of the SM Higgs boson, respectively, modify the kinematic distributions in off-shell ggF Higgs production. Notice that~(\ref{eq:rescale}) leaves the 
total Higgs production cross sections in all channels unchanged compared to  their SM values. This is however not true for  the off-shell Higgs cross sections that are essentially independent of $\Gamma_h^{\rm SM}$ and are thus modified if the Higgs couplings $g_{h XX}^{\rm SM}$ are rescaled as in~(\ref{eq:rescale}). By measuring the total number of off-shell Higgs events one can therefore place indirect limits on the total width of the Higgs boson~\cite{Kauer:2012hd,Caola:2013yja,Campbell:2013una,CMS:2014quz,ATLAS:2015cuo,ATL-PHYS-PUB-2015-024,CMS-PAS-FTR-18-011,ATLAS:2018jym,CMS:2019ekd,ATLAS:2019qet}.

\begin{figure}[!t]
\begin{center}
\includegraphics[width=0.475\textwidth]{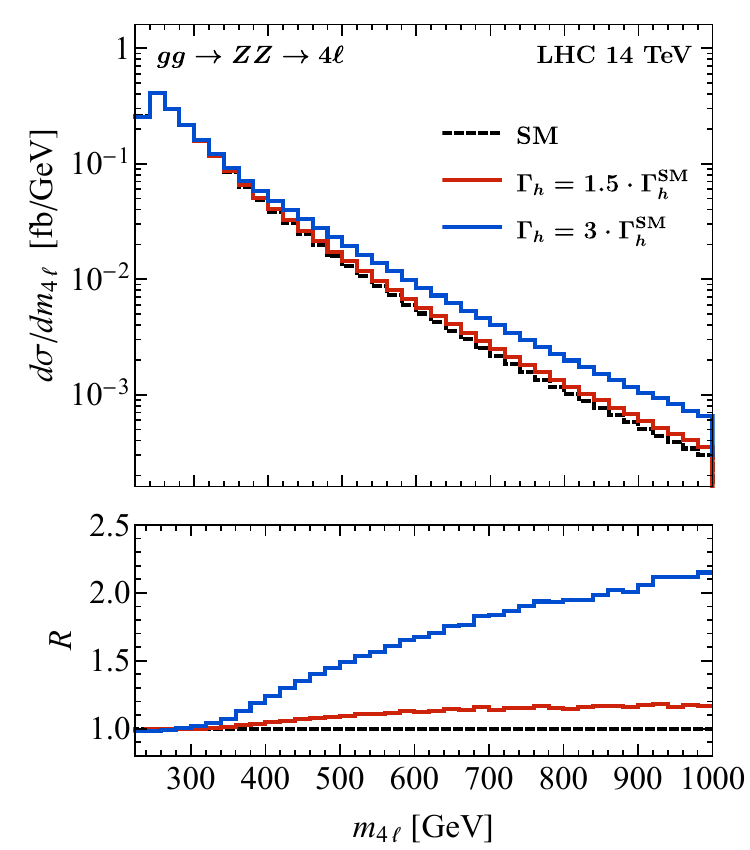} \quad 
\includegraphics[width=0.475\textwidth]{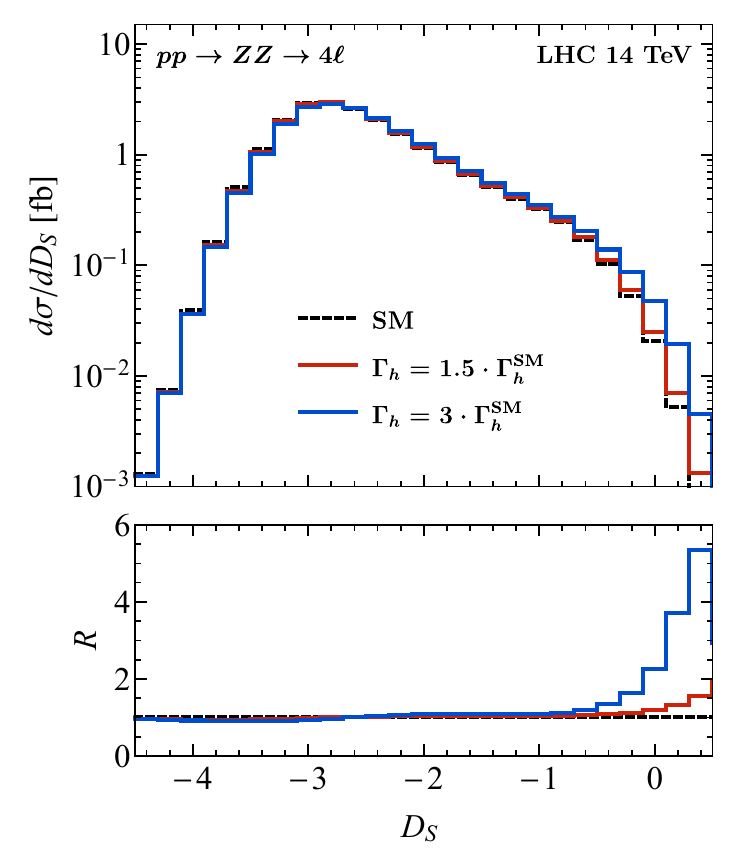}
\vspace{2mm} 
\caption{\label{fig:higgswidth} Left: $m_{4 \ell}$ distributions for the $gg$-initiated contributions in the SM~(dashed black), for $\Gamma_h = 1.5 \cdot  \Gamma_h^{\rm SM}$~(solid red) and for $\Gamma_h = 3 \cdot  \Gamma_h^{\rm SM}$~(solid blue). Right: $D_S$~distributions for the $pp \to Z Z \to 4 \ell$  process in the SM (dashed black), for $\Gamma_h = 1.5 \hspace{0.25mm} \cdot  \hspace{0.25mm}  \Gamma_h^{\rm SM}$~(solid red) and for $\Gamma_h = 3  \hspace{0.25mm} \cdot   \hspace{0.25mm} \Gamma_h^{\rm SM}$~(solid blue). All distributions are obtained using~(\ref{eq:rescale}), are LO~QCD accurate and assume $pp$ collisions at $\sqrt{s} = 14 \, {\rm TeV}$. The lower panels show the ratios between the BSM  and   SM predictions.    } 
\end{center}
\end{figure}

In Figure~\ref{fig:higgswidth} we show  our results for  the $m_{4 \ell}$ distributions in the $gg \to Z Z \to 4 \ell$ channel~(left) and the $D_S$ spectrum of $pp \to Z Z \to 4 \ell$~(right) for two different rescalings~(\ref{eq:rescale}). The choice $\xi = 3$ and $\xi =1.5$ thereby corresponds approximately to the present LHC~Run~2~\cite{ATLAS:2018jym,CMS:2019ekd} and the projected HL-LHC~\cite{ATL-PHYS-PUB-2015-024,CMS-PAS-FTR-18-011} sensitivity, respectively. From the left plot one sees that  compared to the SM the BSM predictions have larger off-shell Higgs cross sections with the relative difference between the spectra growing roughly linearly with~$m_{4 \ell}$. Notice that the observed shape changes are qualitatively different from  the  relative modifications  that occur in the case of the ${\cal O} (\lambda)$ corrections associated to insertions of the operator~$\mathcal{O}_6$ as shown on the right-hand side in Figure~\ref{fig:m4lplot}. From the right plot in Figure~\ref{fig:higgswidth} one furthermore observes that compared to the SM the BSM distributions of the ME-based discriminant are enhanced for $D_S \gtrsim -1$.   Since they do not feature the enhancements for $D_S\lesssim -3.5$, the shown $D_S$ spectra are hence distinct from the distributions that are displayed on the right in Figure~\ref{fig:DSplotpp}, which correspond to the spectra resulting from insertions of the SMEFT operator ${\cal O}_6$. Notice that in contrast to the ${\cal O} (\lambda)$ corrections, the effects of~(\ref{eq:rescale}) lead solely to enhancements in the tail of the $m_{4 \ell}$ distribution. In this case  extra care is required in estimating the systematic uncertainties, because the NLO QCD corrections to the $gg$-induced channel included approximately by means of~(\ref{eq:procedure}) implicitly assume an asymptotic expansion in the top-quark mass (cf.~\cite{Alioli:2021wpn,Buonocore:2021fnj}) of the relevant  two-loop $gg \to ZZ$ amplitudes.  This expansion fails above the top-quark threshold,~i.e.~for four-lepton invariant masses~$m_{4\ell}>2m_t$, which introduces compared to the case discussed in~Section~\ref{sec:QCD} an additional systematic uncertainty.  To account for  this issue, we instead of $\pm 15\%$ assume an enlarged  total theoretical uncertainty of $\pm 25\%$. Employing this uncertainty estimate and performing a shape-fit  to the $D_S$ spectrum following the procedure outlined  in Section~\ref{sec:c6chfit}, we obtain for  $3 \, {\rm ab}^{-1}$ of HL-LHC data
the 95\%~CL bound $\Gamma_h < 1.49 \hspace{.25mm} \cdot \hspace{.25mm} \Gamma_h^{\rm SM}$. This finding is in line with the limits reported in~\cite{ATL-PHYS-PUB-2015-024,CMS-PAS-FTR-18-011} which validates the used fitting~approach.

\section{Additional HL-LHC projections}
\label{app:HLLHCprojections}

A crucial ingredient in the shape fit to the~$D_S$ distribution described in Section~\ref{sec:c6chfit}  are the systematic uncertainties~$\sigma_{b_i}$ on the background as parametrised by the parameters~$ \Delta_i$  in~(\ref{eq:sbsigma}).  In this appendix we present results for two additional  more aggressive assumptions about the systematic uncertainties entering the HL-LHC off-shell Higgs  analysis. Specifically, we   will employ the two different choices~$\Delta_i = 0.08$ and~$\Delta_i  = 0.04$  of bin-independent systematic uncertainties. These choices can be motivated by recalling that the systematic uncertainties that ATLAS quotes in the HL-LHC study~\cite{ATL-PHYS-PUB-2018-054} for the on-shell~$gg \to h \to ZZ$ signal strength amount to~$5.0\%$ and~$3.9\%$ in the baseline scenario S1 and S2 for the expected total  systematic uncertainties. The corresponding systematic uncertainties quoted in the CMS work~\cite{CMS-PAS-FTR-18-011} are~$7.3\%$ and~$4.1\%$. Since the ${\cal O} (\lambda)$  corrections to~$D_S$ considered in this work are associated to kinematic configurations with~$m_{4 \ell}$ not far above~$2 m_Z$, it seems not unreasonable that  theoretical predictions of the~$D_S$ spectra can reach an accuracy that is very similar to the systematics that is expected to be achievable at the HL-LHC in the case of on-shell~$gg \to h  \to ZZ$ production.

\begin{figure}[!t]
\begin{center}
\includegraphics[width=0.475\textwidth]{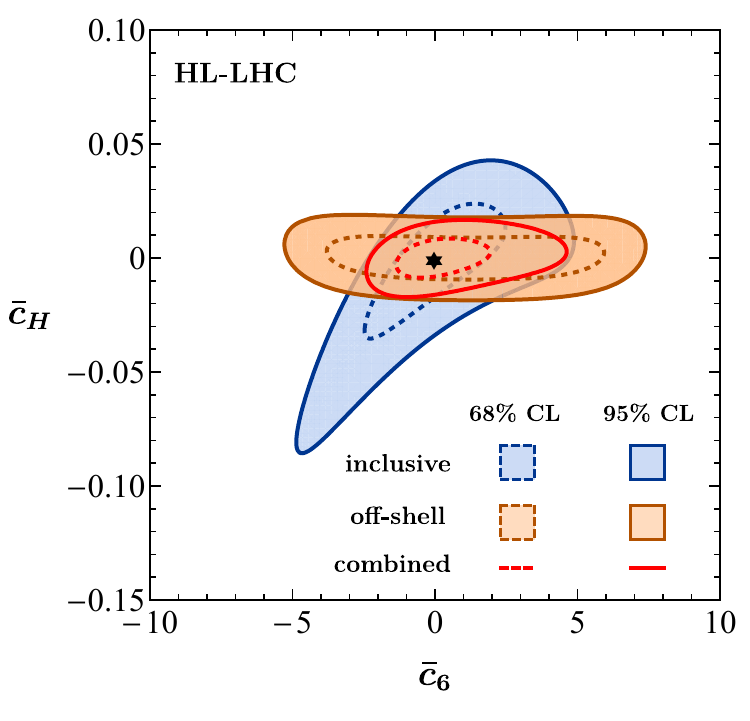} \quad 
\includegraphics[width=0.475\textwidth]{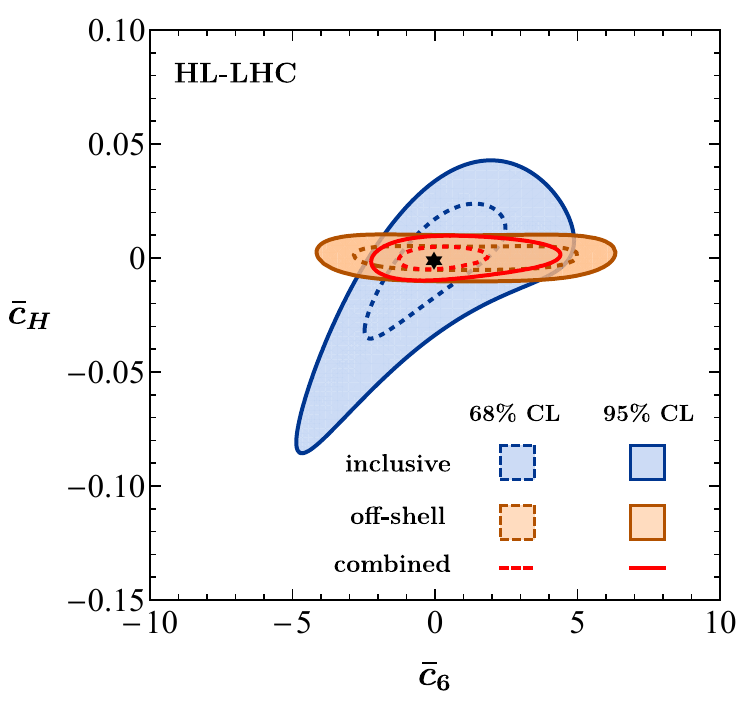}
\vspace{2mm} 
\caption{\label{fig:projections} Projected 68\% and 95\%~CL constraints in the $\bar c_6 \hspace{0.25mm}$--$ \hspace{0.25mm} \bar c_H$ plane for the  HL-LHC assuming $3 \, {\rm ab}^{-1}$ of data collected at $\sqrt{s} = 14 \, {\rm TeV}$. The constraints from inclusive single-Higgs probes (blue regions) are compared to the off-shell Higgs constraints (orange regions). The left (right)  off-shell Higgs constraints  employ a bin-independent systematic uncertainty of $\Delta_i = 0.08$ ($\Delta_i = 0.04$) in the shape fit to the~$D_S$ distribution. The combinations of all constraints are also shown as red contours. The black stars represent the SM point. For~additional details see the main text. }  
\end{center}
\end{figure} 

In the left (right) panel of Figure~\ref{fig:projections} we show the projected 68\% and 95\%~CL HL-LHC constraints in the $\bar c_6 \hspace{0.25mm}$--$ \hspace{0.25mm} \bar c_H$ plane assuming $\Delta_i = 0.08$ ($\Delta_i = 0.04$). The constraints from inclusive single-Higgs probes~(blue regions) are compared to the off-shell Higgs constraints~(orange regions). Their combinations (red contours) are also displayed. From a combined analysis of inclusive single-Higgs and off-shell Higgs probes, we find for $\bar c_H =0$ the  following 95\%~CL limits 
\begin{equation} \label{eq:c6boundsallproj}
\begin{split}
\bar{c}_6 & \in [-2.3, 4.5] \,, \;\; (\text{HL-LHC}, \, \Delta_i = 0.08) \,, \\[2mm]
\bar{c}_6 & \in [-2.2, 4.3] \,, \;\;   (\text{HL-LHC}, \, \Delta_i = 0.04) \,,
\end{split}
\end{equation}
while for $\bar c_6 =0$ we obtain 
\begin{equation} \label{eq:cHboundsallproj}
\begin{split}
\bar{c}_H & \in [-1.6, 1.6] \cdot 10^{-2} \,, \;\; (\text{HL-LHC},\,  \Delta_i = 0.08)  \,, \\[2mm]
\bar{c}_H & \in [-1.0, 1.0]  \cdot 10^{-2}  \,, \;\; (\text{HL-LHC}, \Delta_i = 0.04) \,.
\end{split}
\end{equation}
By comparing the HL-LHC limits given in (\ref{eq:c6boundsall}) and (\ref{eq:cHboundsall}) to the above results, one observes that a reduction of systematic uncertainties has only a minor impact in the case of $\bar c_6$, while it has a noticeable impact on the resulting bounds on $\bar c_H$. The limits~(\ref{eq:c6boundsallproj}) and~(\ref{eq:cHboundsallproj}) can also be translated into constraints on the modifications of the trilinear Higgs coupling as parameterised by~(\ref{eq:c3}). The corresponding 95\%~CL ranges read $c_3  \in [-1.4, 5.6]$ and $c_3  \in [-1.2, 5.4]$, respectively, which again represent only minor improvements compared to the HL-LHC bound derived in the main body of this article.


\providecommand{\href}[2]{#2}\begingroup\raggedright\endgroup

\end{document}